\documentclass[twocolumn,aps,pra,preprintnumbers,bibnotes10pt,superscriptaddress,amsmath,amssymb]{revtex4}
\usepackage{graphicx,subfigure}
\usepackage{float}
\usepackage{color}
\usepackage{epstopdf}
\usepackage{amsmath}
\usepackage{amsthm}
\usepackage{amssymb}
\usepackage{amsfonts}
\usepackage{graphicx}
\usepackage{txfonts}
\usepackage{ulem}
\usepackage{mathtools}
\usepackage{bm}

\newcommand{\bra}[1]{\ensuremath{\langle#1|}}
\newcommand{\ket}[1]{\ensuremath{|#1\rangle}}

\newcommand{\be}{\begin{equation}}
\newcommand{\ee}{\end{equation}}
\newcommand{\beq}{\begin{eqnarray}}
\newcommand{\eeq}{\end{eqnarray}}

\DeclarePairedDelimiter\floor{\lfloor}{\rfloor}

\newcommand{\mean}[1]{\ensuremath{\big\langle #1 \big\rangle}}

\begin{document}
\title{Improved absolute clock stability  by the joint interrogation of two atomic states}
\author{Weidong Li}
\affiliation{Institute of Theoretical Physics and Department of Physics, Shanxi University, Taiyuan 030006, China}

\author{Shuyuan Wu}
\affiliation{QSTAR, INO-CNR and LENS, Largo Enrico Fermi 2, 50125 Firenze, Italy}

\author{Augusto Smerzi}
\affiliation{Institute of Theoretical Physics and Department of Physics, Shanxi University, Taiyuan 030006, China}
\affiliation{QSTAR, INO-CNR and LENS, Largo Enrico Fermi 2, 50125 Firenze, Italy}

\author{Luca Pezz\`e}
\affiliation{Institute of Theoretical Physics and Department of Physics, Shanxi University, Taiyuan 030006, China}
\affiliation{QSTAR, INO-CNR and LENS, Largo Enrico Fermi 2, 50125 Firenze, Italy}

\begin{abstract}
Improving the clock stability is of fundamental importance for the development of quantum-enhanced metrology.
One of the main limitations arises from the randomly-fluctuating local oscillator (LO) frequency, which introduces ``phase slips'' for long interrogation times and hence failure of the frequency-feedback loop.
Here we propose a strategy to improve the stability of atomic clocks by interrogating two out-of-phase state sharing the same LO.
While standard Ramsey interrogation can only determine phases unambiguously in the interval $[-\pi/2,\pi/2]$, the joint interrogation allows for an extension to $[-\pi,\pi]$, resulting in a relaxed restriction of the Ramsey time and improvement of absolute clock stability.
Theoretical predictions are supported by ab-initio numerical simulation for white and correlated LO noise.
While our basic protocol uses uncorrelated atoms, 
we have further extended it to include spin-squeezing 
and further improving the scaling of clock stability with the number of atoms. 
Our protocol can be readily tested in current state-of-the-art experiments. 
\end{abstract}

\maketitle

\section{Introduction}
\begin{figure}[t!]
\includegraphics[width=\columnwidth]{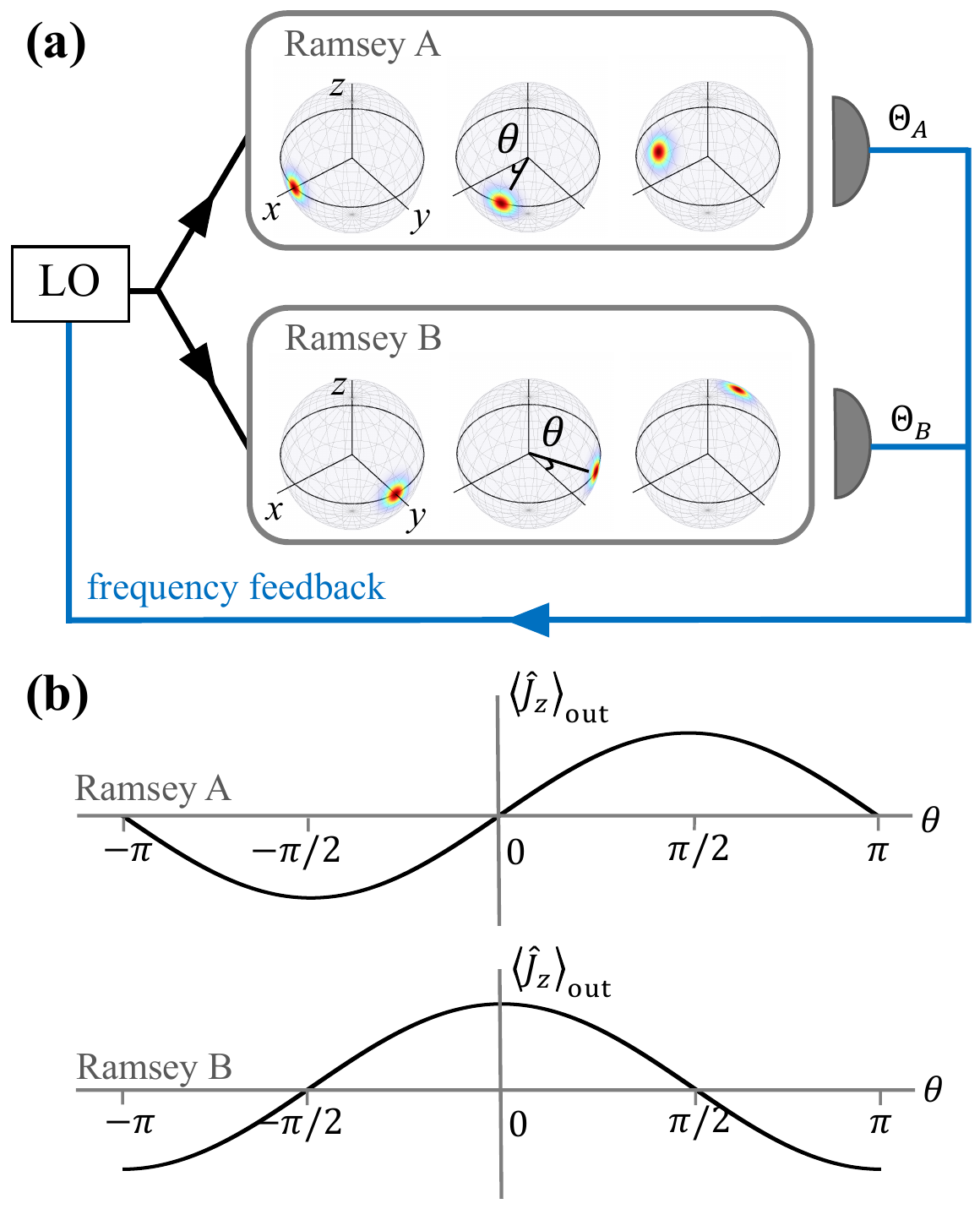}
\caption{(a) Scheme of the joint interrogation method. 
Two Ramsey interferometers share the same LO and atomic transition such that
the accumulated phase $\theta$ during the common Ramsey interrogation time is the same for both states.
The probes are coherent spin states with mean spin direction pointing along different axis in the Bloch sphere 
($x$ axis for the interferometer A and $y$ axis for the interferometer B).
The inset shows Husimi distributions of the state during the different interferometer operations.
(b) Ramsey signal $\mean{J_z(\theta)}_{\rm out}$ as a function of $\theta$ for the two interferometers.
Combining the two out-of-phase signals (see text) it is possible to obtain an unbiased estimate 
of $\theta$ in the full $[-\pi, \pi]$ interval.} 
\label{Figure1}
\end{figure}

The basic working principle of a passive atomic clock~\cite{Vanier1992, Wynands, Riehle2004, PoliNC2013, LudlowRMP2015} is to stabilize the frequency of a local oscillator (LO) 
to an atomic resonance $\omega_0$. 
Using the Ramsey interferometer method~\cite{Ramsey1956}, the unavoidable frequency fluctuations of the LO accumulate during an interrogation time and 
result to an overall rotation of the collective pseudo-spin of $N$ two-level atoms by a stochastic angle $\theta$. 
This angle is estimated by measuring the population imbalance of the two clock levels, which is a sinusoidal function of $\theta$ 
and can be inverted only in a restricted region where the output signal is monotonic, also indicated as inversion region. 
The phase estimate is converted in a time-averaged frequency estimate that is used to steer the LO frequency toward $\omega_0$.
The frequency stability can be improved by increasing the interrogation time as far as
$\theta$ remains within the inversion region. 
The stochastic occurrence of a value of $\theta$ outside the inversion region is generally indicated as 
 ``phase slip'' (or ``fringe hop'') and prevents the unbiased correction of the LO frequency. 
Within conventional phase estimation methods, phase slips occur when $\vert \theta \vert \geq \pi/2$, see Fig.~\ref{Figure1} and details below.
In current atomic clocks -- especially those exploiting ultrastable lasers~\cite{McGrew2018, Bloom2014, TakamotoNat2005,JiangNPhoton2011,Thorpe2011,Kessler2012,Cole2013,Nemitz2016} -- the LO decoherence dominates over the atomic decoherence and thus sets the crucial limitation to clock stability.

Methods to avoid phase-slips and thus extend the interrogation time have an immediate practical relevance and 
are thus attracting increasing interest in the literature~\cite{Rosenband2013, HumePRA2016, BorregaardPRL2013_b, ShigaNJP2012, ShigaNJP2014, KohlhaasPRX2015, KaubrueggerARXIV}.
Some proposals~\cite{Rosenband2013, HumePRA2016} have considered the simultaneous use of multiple atomic ensembles
characterized by different transition frequencies (e.g. an optical lattice clock and a single-ion clock), 
which are phase locked via a frequency comb.
This method allows to extend the interrogation time of the ensemble characterized by the higher atomic transition frequency. 
In contrast, Ref.~\cite{Rosenband2013, BorregaardPRL2013_b} considered ensembles having the same transition frequency but probed for different times.  
In this case, it is possible to extend the interrogation of the ensemble probed for the shorter time.
Yet, both these methods do not allow to extend the interrogation time of the ensemble characterized by the smaller transition frequency or the ensemble probed for the longer time.
Another possibility is to phase-lock -- via successive quantum non-demolition measurements -- the LO to the atomic state and thus increase the 
Ramsey interrogation time while avoiding phase slips~\cite{ShigaNJP2012, ShigaNJP2014, KohlhaasPRX2015, ColangeloNATURE2017}.
Finally, Ref.~\cite{KaubrueggerARXIV} has recently discussed a variational optimization 
realizing measurements similar to Pegg-Barnett phase operators~\cite{PeggEPL1998, BuzekPRL1999}, which have eigenstates with well defined phases. 

In this manuscript, we propose a strategy to improve the frequency stability by interrogating two coherent 
spin states that share the same LO, same interrogation time and the same atomic transition, see Fig.~\ref{Figure1}.
The two interferometers are characterized by Ramsey fringes that are dephased by $\pi/2$.
By combining the independent phase estimates obtained from the two interferometers, it is possible to extend the inversion region 
from $[-\pi/2,\pi/2]$ to $[-\pi,\pi]$: this results in an increase of the optimal Ramsey time and therefore also an increase of the absolute stability of the clock.
Taking into account basic models of LO decoherence, we predict an improvement of absolute stability by a factor four (two) for the case of white (flicker) LO noise, when compared to the case of single-ensemble interrogation.
Our protocol, differently from Refs.~\cite{ShigaNJP2012, ShigaNJP2014, KohlhaasPRX2015}, 
does not require quantum non-demolition measurements and, differently from Refs.~\cite{Rosenband2013, HumePRA2016, BorregaardPRL2013_b}, it 
allows to extend the overall absolute stability.
Our methods use only classical resources, namely,
unentangled atoms, and can be generalized to include spin-squeezed atomic ensembles to overcome the standard quantum limit. 

The paper is organized as follows.
In Sec.~\ref{Sec2}, we introduce the 
Ramsey interferometer of Fig.~\ref{Figure1} and discuss in details its phase sensitivity.
In Sec.~\ref{Sec3} we review basic concepts of atomic clocks.
In particular, we focus on the detailed calculation of the optimal (Ramsey) interrogation time by carefully evaluating the bending point of the Allan variance
(defining the absolute stability of the clock).
Our careful methods to calculate the Allan variance avoid numerical instabilities that characterize standard protocols in the presence of phase slips.  
Furthermore, in the case of white (correlated) LO noise, we provide analytical (semi-analytical) predictions that are found in full agreement with the results of ab-initio MonteCarlo simulations. 
In Sec.~\ref{Sec4}, we analyze the impact of possible experimental imperfections in the state preparation, fluctuations of the number of particles and dead times.
Finally, in Sec.~\ref{Sec5}, we exploit the joint-Ramsey protocol in connection to a recent proposal of a hybrid quantum-classical clock using 
coherent- and squeezed-spin states~\cite{PezzePRL2020}.

%%%%%%%%
%% Figure 1
%%%%%%%%
%%%%%%%%
%%%%%%%%
%%%%%%%%

\section{Ramsey interferometry and phase sensitivity}
\label{Sec2}

We consider $N$ atoms, each modeled as a pseudospin-$1/2$ particle and
introduce collective spin operators 
$\hat{J}_{x,y,z} = \sum_{j=1}^N \hat{\sigma}_{x,y,z}^{(j)}/2$, where $\hat{\sigma}_x = \ket{\uparrow} \bra{\downarrow} + \ket{\downarrow} \bra{\uparrow}$, 
$\hat{\sigma}_y = i(\ket{\uparrow} \bra{\downarrow} - \ket{\downarrow} \bra{\uparrow})$ and 
$\hat{\sigma}_z = \ket{\uparrow} \bra{\uparrow} - \ket{\downarrow} \bra{\downarrow}$ are Pauli operators.
In particular $\hat{J}_z$ counts the relative number of particles
between the two clock levels $\ket{\uparrow}$ and $\ket{\downarrow}$. 

During the free evolution following the state preparation, the collective atomic pseudospin processes around the $z$ axis by an angle $\theta$, which we specify later.
The Ramsey sequence terminates with a $\pi/2$ rotation around the $x$ axis that converts the rotation angle $\theta$ into a population difference.
We assume that the duration of the $\pi/2$ pulse is short enough so to neglect fluctuations of the LO leading to imperfections in the pulse rotation angle.
In the noiseless case, the inteferometer sequence for both states in Fig.~\ref{Figure1}(a) is described by the unitary transformation
$\hat{U}(\theta)=\text{e}^{-i\frac{\pi}{2}\hat{J}_x}\text{e}^{-i \theta \hat{J}_z}$.
The relative number of particle operator transforms according to 
$\hat{J}_z(\theta)=\hat{U}^{\dagger}(\theta)\hat{J}_z\hat{U}(\theta)=\hat{J}_y\text{cos} \theta + \hat{J}_x\text{sin} \theta$.
A measurement of $\hat{J}_z(\theta)$ provides an estimate of $\theta$, depending on the interferometer scheme (single or joint, that we describe below).

\subsection{Single-Ramsey interferometer}
\label{Sec.SingleRamsey}

In the Ramsey interferometer A of Fig.~\ref{Figure1}(a), the atoms are prepared in the coherent-spin state
\be \label{CSS1}
\ket{\psi_A} = \bigg( \frac{\ket{\uparrow} +\ket{\downarrow}}{\sqrt{2}} \bigg)^{\otimes N},
\ee
with mean spin direction pointing along the $x$ axis,
$\mean{\hat{J}_x}_{\rm in}  = \bra{\psi_A} \hat{J}_x \ket{\psi_A} = N/2$.
Being $\mean{\hat{J}_y}_{\rm in} = 0$, 
the Ramsey signal is given by
\begin{equation} \label{Ramsey1}
\mean{ \hat{J}_z (\theta)}_{\rm out}=\frac{N}{2}\sin \theta,
\end{equation}
as shown in Fig.~\ref{Figure1}(b).
Based on a single measurement of $\hat{J}_z$ with result $\mu_A = (N_\uparrow - N_{\downarrow})/2$, where $N_\uparrow$ ($N_\downarrow$)
is the number of particles in $\ket{\uparrow}$ ($\ket{\downarrow}$),
we obtain the estimate 
\be \label{est}
\Theta_A(\mu_A)={\rm arcsin} \frac{2 \mu_A}{N}
\ee 
of $\theta$.
Notice that $\Theta_A(\mu_A)$ is a value in the interval $[-\pi/2, \pi/2]$
where Eq.~(\ref{Ramsey1}) is monotonic, see Fig.~\ref{Figure1}(b).
This interval will be indicated as inversion region in the following. 

The estimator Eq. (\ref{est}) is a random variable with statistical mean value
$\bar{\Theta}_A(\theta) = \mathcal{E}_{\mu_A \vert \theta}[\Theta_A(\mu_A)]$, where 
$\mathcal{E}_{\mu_A \vert \theta}[...] = \sum_{\mu_A}P(\mu_A  \vert \theta) ...$ indicates the statistical average 
over random measurement results obtained for a fixed value of $\theta$, 
and $P(\mu_A \vert \theta)$ is the corresponding conditional probability to obtain the result $\mu_A$.
The fluctuations of the estimator are quantified by the variance 
$\big( \Delta  \Theta_A(\theta) \big)^2= \mathcal{E}_{\mu_A \vert \theta}\big[\big( \bar{\Theta}_A(\theta)  - \Theta_A(\mu_A) \big)^2\big]$
that can be well approximated by the error propagation formula
\be \label{Dtheta}
\big( \Delta  \Theta_A(\theta) \big)^2 \approx \frac{ (\Delta \hat{J}_z(\theta))_{\rm out}^2}{\big( d \mean{\hat{J}_z(\theta)}_{\rm out}/d\theta \big)^2} = \frac{1}{N_{t}},
\ee
where $N_{t} = N$ is the total number of particles used in the interferometer.
The right-hand side of the above equation is obtained using $\big(\Delta \hat{J}_z(\theta)\big)^2_{\rm out}  = \mean{\hat{J}_z^2}_{\rm out} - \mean{\hat{J}_z}^2_{\rm out} = 
(N/4) \cos^2 \theta$, which is due to $(\Delta \hat{J}_x)^2_{\rm in} = 0$, $\mean{\hat{J}_x \hat{J}_z + \hat{J}_z \hat{J}_x}_{\rm in} = 0$ and $(\Delta \hat{J}_z)^2_{\rm in} = N/4$, for the coherent state Eq. (\ref{CSS1}).

%%%%%%%%%%%%%%%%%%%%%%%%%%%%%%%%%%%%%%%%%%%%%%%
%% Figure 2
%%%%%%%%%%%%%%%%%%%%%%%%%%%%%%%%%%%%%%%%%%%%%%%
\begin{figure}[t!]
\includegraphics[width=\columnwidth]{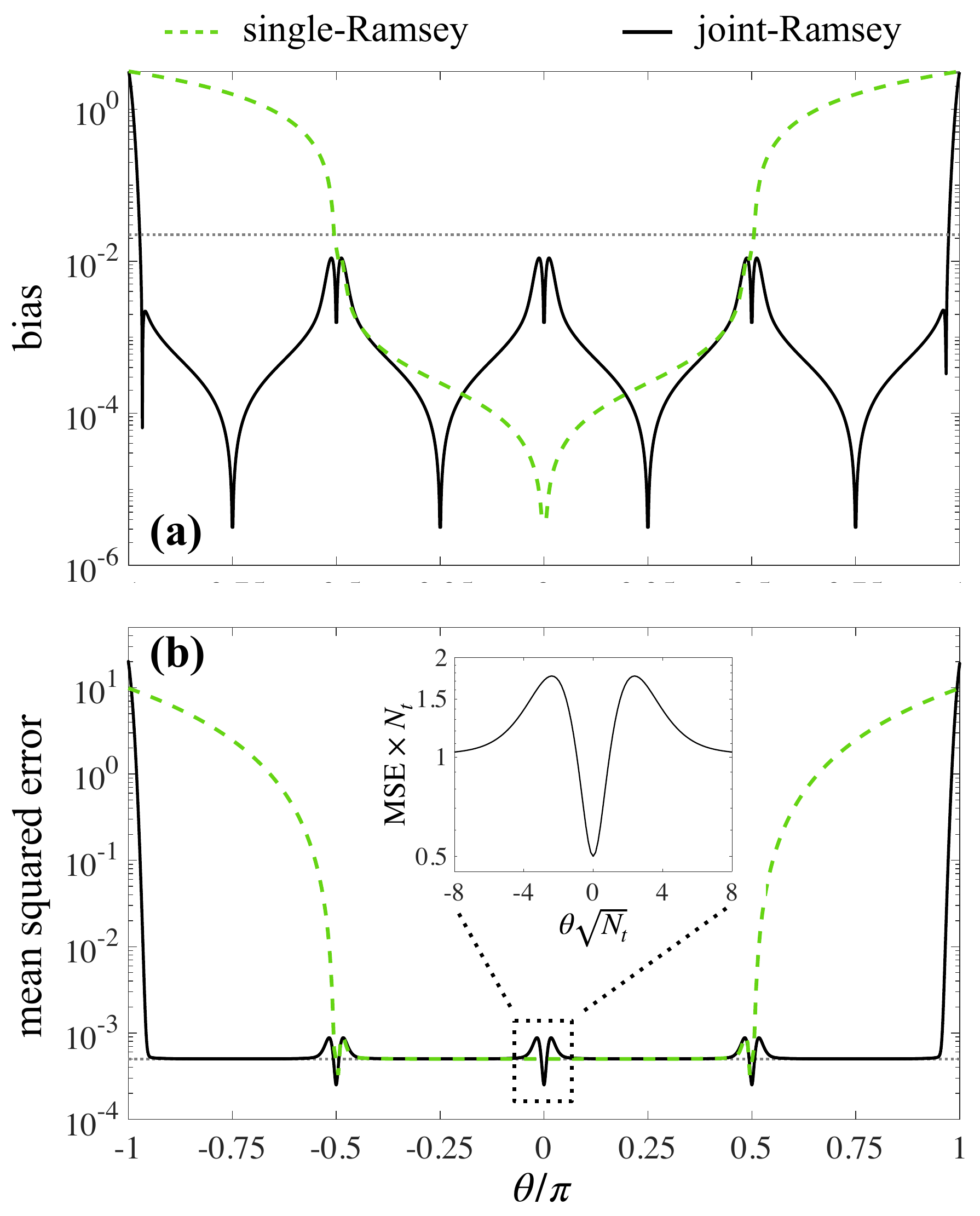}
\caption{
(a) Estimator bias as a function of $\theta$.
The green dashed line is $\vert \theta - \bar{\Theta}_A(\theta)\vert$ for the single-Ramsey interferometer with Eq.~(\ref{CSS1}) as input, while
the black solid line is $\vert \theta - \bar{\Theta}_{AB}(\theta)\vert$ for the joint-Ramsey scheme.
The horizontal dotted line is $1/\sqrt{N_t}$.
(b) Mean squared error as a function of $\theta$.
The green dashed (black solid) line is obtained the single- (joint-) Ramsey interferometer. 
The horizontal dotted line is $1/N_t$.
The inset is a zoom showing the mean squared error multiplied by $N_t$ as a function of $\theta \sqrt{N_t}$ around $\theta=0$.
In both panels, the single and joint protocols are compared by fixing the total number of particles $N_t = 2000$, that is 
$N=N_t$ for the single-Ramsey interferometer and $N=N_t/2$ in each state of the joint-Ramsey scheme.}
\label{Figure2}
\end{figure}
%%%%%%%%%%%%%%%%%%%%%%%%%%%%%%%%%%%%%%%%%%%%%%%
%%%%%%%%%%%%%%%%%%%%%%%%%%%%%%%%%%%%%%%%%%%%%%%
%%%%%%%%%%%%%%%%%%%%%%%%%%%%%%%%%%%%%%%%%%%%%%% 

In Fig.~\ref{Figure2}(a) we show the estimator bias $\vert \theta - \bar{\Theta}_A(\theta)\vert$ (green dashed line)
obtained from an exact numerical calculation.
For $\vert \theta \vert \lesssim \pi/2$, we have $\vert \theta - \bar{\Theta}_A(\theta) \vert \ll \Delta  \Theta_A(\theta)$:
the distance between the true value $\theta$ and the mean value of the estimator is smaller than the statistical fluctuations of the estimator itself. 
Instead, for $ \vert \theta \vert \geq  \pi/2$, outside the inversion region, the estimate (\ref{est}) is characterized by a finite bias 
$\vert \theta - \bar{\Theta}_A(\theta)  \vert 
 \approx  2 \vert \theta \vert  - \pi$.
 In Fig.~\ref{Figure2}(b) we show the mean squared error of $\Theta_A$,
 \be \label{Dtheta0}
\mathcal{E}_{\mu_A \vert \theta} \big[ \big( \Theta_A(\mu_A) - \theta \big)^2 \big]= \big( \Delta  \Theta_A(\theta) \big)^2 + \big(\theta - \bar{\Theta}_A(\theta) \big)^2.
 \ee
For $\vert \theta \vert \lesssim \pi/2$ we have $\mathcal{E}_{\mu_A \vert \theta} \big[ \big( \Theta_A(\mu_A) - \theta \big)^2 \big] \approx  \big( \Delta  \Theta_A(\theta) \big)^2 \approx 1/N_t$,
as predicted by Eq.~(\ref{Dtheta}).
Notice that a characteristic property of Eq.~(\ref{Dtheta}) is the independence from $\theta$, which is confirmed by the exact numerical 
calculations.
For $\vert \theta \vert \geq  \pi/2$, the bias dominates and Eq.~(\ref{Dtheta0}) is well approximated by $1/N_t + (2 \vert \theta \vert  - \pi)^2$. 

\subsection{Joint-Ramsey interferometer}
\label{Sec.joint}

The joint interferometer scheme is shown in Fig.~\ref{Figure1}(a).
The first interferometer [indicated as Ramsey A in Fig.~\ref{Figure1}(a)] is a standard one, as discussed above.
The second interferometer [Ramsey B] differs from the first one by the initial state, which is given by 
\be \label{CSS2}
\ket{\psi_B} = \bigg(\frac{\ket{\uparrow} + i \ket{\downarrow}}{\sqrt{2}}\bigg)^{\otimes N}
\ee
and has a mean spin direction pointing along the $y$ axis, $\mean{\hat{J}_y}_{\rm in}  = \bra{\psi_B} \hat{J}_y \ket{\psi_B} = N/2$.
The Ramsey signal for the interferometer B is 
\begin{equation} \label{JzRamsey2}
  \mean{ \hat{J}_z(\theta) }_{\rm out}=\frac{N}{2}\text{cos}\theta,
\end{equation}
with fluctuations
$\big( \Delta \hat{J}_z(\theta) \big)^2_{\rm out}  = (N/4) \sin^2 \theta$.
A measurement of $\hat{J}_z$ with result $\mu_B$, leads to an estimate 
\be \label{est2}
\Theta_B(\mu_B)={\rm arccos} \frac{2 \mu_B}{N},
\ee
obtained by inverting Eq.~(\ref{JzRamsey2}).
Notice that Eq.~(\ref{JzRamsey2}) is characterized by two inversion regions where the function of $\theta$ is monotonic: 
$[-\pi,0]$ and $[0, \pi]$, see Fig.~\ref{Figure1}(b).
Equation~(\ref{JzRamsey2}) is dephased by $\pi/2$ with respect to Eq.~(\ref{Ramsey1}) due to the different initial state.
The total number of particles used in the 
joint interferometer is $N_t=2N$, where $N$ is the number of particles in each state (\ref{CSS1}) and (\ref{CSS2}).

The central idea of this manuscript is to combine the two estimates $\Theta_A(\mu_A)$ and $\Theta_B(\mu_B)$ to obtain  a joint estimate $\Theta_{AB}(\mu_A, \mu_B)$ of $\theta$.
Specifically,  we define 
\be \label{thest12}
\Theta_{AB}(\mu_A, \mu_B) = 
\begin{cases}
\frac{\Theta_A(\mu_A)+\Theta_B(\mu_B)}{2} & \text{if $\mu_A > 0$ and $\mu_B > 0$}, \\
\frac{\pi - \Theta_A(\mu_A)+\Theta_B(\mu_B)}{2} & \text{if $\mu_A > 0$ and $\mu_B < 0$}, \\
\frac{\Theta_A(\mu_A) - \Theta_B(\mu_B)}{2} & \text{if $\mu_A < 0$ and $\mu_B > 0$}, \\
\frac{-\pi - \Theta_A(\mu_A) - \Theta_B(\mu_B)}{2} & \text{if $\mu_A < 0$ and $\mu_B > 0$},
\end{cases}  
\ee
and, when the measurement results are $\mu_{A}=0$ or $\mu_B=0$, 
\be \label{thest12b}
\Theta_{AB}(\mu_A, \mu_B) = 
\begin{cases}
0 & \text{if $\mu_A = 0$ and $\mu_B > 0$}, \\
\pi & \text{if $\mu_A = 0$ and $\mu_B < 0$}, \\
\pi/2 & \text{if $\mu_A > 0$ and $\mu_B = 0$}, \\
-\pi/2 & \text{if $\mu_A < 0$ and $\mu_B = 0$}. \\
\end{cases}
\ee
For instance, if $\mu_A \geq 0$ and $\mu_B \geq 0$, the true phase is most likely in the region $[0,\pi/2]$
where $\mean{\hat{J}_z}_{\rm out} \geq 0$ for both interferometers, see Fig.~\ref{Figure1}(b).
In this case, the joint estimate $\Theta_{AB}(\mu_a, \mu_B)$ is thus simply chosen as the sum of 
$\Theta_A$ and $\Theta_B$, divided by two.
Conversely, if $\mu_A \leq 0$ and $\mu_B \geq 0$, the true phase is most likely in the region $[-\pi/2,0]$
where the Ramsey signal $\mean{\hat{J}_z}_{\rm out}$ is negative for the interferometer A and positive for the interferometer B, see Fig.~\ref{Figure1}(b). 
In this case, $\Theta_{AB}(\mu_a, \mu_B)$ is chosen as the difference between $\Theta_A$ and $\Theta_B$ divided by two, and so on.
This explains the choice of linear combination of 
$\Theta_A$ and $\Theta_B$ in Eq.~(\ref{thest12}).

In Fig.~\ref{Figure2}(a) we plot the bias $\vert \theta  - \bar{\Theta}_{AB}(\theta) \vert$ as a function of $\theta$, where $\bar{\Theta}_{AB}(\theta) =\mathcal{E}_{\mu_A, \mu_B \vert \theta}[\Theta_{AB}(\mu_A, \mu_B)]$
is the statistical mean value of the estimator (\ref{thest12}), 
and
$\mathcal{E}_{\mu_A, \mu_B \vert \theta}[...] = 
\sum_{\mu_A, \mu_B} P(\mu_A \vert \theta) P(\mu_B\vert \theta) ...$
indicates statistical averaging over random measurement results $\mu_A$ and $\mu_B$.
The bias is essentially negligible, $\vert \theta  - \bar{\Theta}_{AB}(\theta) \vert \ll 1/\sqrt{N_t}$,
in the full $\theta\in [-\pi, \pi]$ interval, except in  
regions of width approximately $1/\sqrt{N}$ close to $\theta=0,\pm\pi/2, \pm \pi$. 
For instance, for $0 \lesssim \theta \lesssim 1/\sqrt{N}$, one would expect to observe results 
$\mu_A \geq 0$ [since $\mean{\hat{J}_z(\theta)}_{\rm out} \geq 0$ for the interferometer A is positive for these values of $\theta$, see Fig.~\ref{Figure1}(b)], but 
results $\mu_A<0$ are also possible due to the finite width of the relative number of particles distribution $P(\mu_A \vert \theta)$.
In this case, according to Eq.~(\ref{thest12}), we combine the two estimates according to $\Theta_{AB}= (\Theta_A-\Theta_B)/2$, thus introducing a bias approximately equal to $2\theta$.
Similar considerations hold close to $\pm \pi/2$.
Close to $\pm \pi$ the bias is much larger than around $0$ and $\pm \pi/2$, see Fig.~\ref{Figure2}(a).
Similarly as above, this bias is related the finite width of the relative number of particles.
For instance, for $\pi - 1/\sqrt{N} \lesssim \theta \lesssim \pi$, 
it may happen that the measurement results give $\mu_A<0$ and $\mu_B<0$ (rather than the expected $\mu_A>0$ and $\mu_B<0$),
thus introducing a large bias of approximately $2\pi$.
This qualitatively explains the increase 
of $\vert \theta  - \bar{\Theta}_{AB}(\theta) \vert$ close to $\pm \pi$, as shown in Fig.~ 
\ref{Figure1}(a).

In Fig.~\ref{Figure2}(b) we plot the mean square error around $\theta$, 
\be \label{fluctuationsAB}
\mathcal{E}_{\mu_A, \mu_B \vert \theta} \big[ \big( \Theta_{AB}(\mu_a, \mu_B) - \theta \big)^2 \big]= \big( \Delta  \Theta_{AB}(\theta) \big)^2 + \big(\theta - \bar{\Theta}_{AB}(\theta) \big)^2,
\ee
where 
$\big( \Delta {\Theta}_{AB}(\theta) \big)^2 = \mathcal{E}_{\mu_A, \mu_B \vert \theta}\big[\big( \bar{\Theta}_{AB}(\theta)  - \Theta_{AB}(\mu_A, \mu_B) \big)^2\big]$
is the statistical variance of the joint estimator.
Neglecting effects associated to the bias predicts 
\be \label{DthetaAB}
\big( \Delta \Theta_{AB} (\theta) \big)^2 = 
\frac{ \big( \Delta \Theta_{A} (\theta) \big)^2 + \big( \Delta \Theta_{B} (\theta)  \big)^2}{4} = 
\frac{1}{N_t}.
\ee
Equation (\ref{DthetaAB}) is numerically verified in the full $\theta\in [-\pi, \pi]$ interval except close 
to $0, \pm \pi/2, \pm \pi$ where the effect of the bias is not negligible.
For instance, at $\theta=0$, we find that 
$\mean{\hat{J}_z}_{\rm out} = N/2$ for the interferometer $B$, 
which implies $\mu_B = N/2$ and $\Theta_B(N/2)=0$ [with vanishing statistical fluctuations since $(\Delta \hat{J}_z)^2_{\rm out} = 0$].
The estimator $\Theta_B$ has  $d\bar{\Theta}_B(\theta)/d \theta \ll 1$
close to $\theta=0$ and it is thus strongly biased.
Therefore, according to Eq.~(\ref{thest12}), $\Theta_{AB}(\mu_A, \mu_B) = \Theta_A(\mu_A)/2$, which implies $(\Delta \Theta_{AB})^2 = (\Delta \Theta_A)^2/2 = 1/(2N_t)$: the variance of the estimator $\Theta_{AB}$ drops by a factor 2 with respect to Eq.~(\ref{DthetaAB}) due to the bias.
The behaviour of $(\Delta \Theta_{AB})^2$ close to $\theta=0$ is shown in the inset of Fig.~\ref{Figure2}(b).
A similar behaviour is observed close to $\theta = \pm \pi/2$, 
where $\mean{\hat{J}_z}_{\rm out} = \pm N/2$ for the interferometer A with vanishing fluctuations.
Close to $\theta=\pm \pi$, the mean square error Eq.~(\ref{fluctuationsAB}) increases substantially due to the strong bias.

\section{Clock operations and figure of merit}
\label{Sec3}

The clock operations are described by introducing three relevant quantities:
i) the atomic transition frequency $\omega_0$ between two atomic levels;
ii) the free running (unlocked) LO frequency $\tilde{\omega}_{\rm LO}(t)$; and
iii) the stabilized LO frequency $\omega_{\rm LO}(t)$ obtained from periodic feedback corrections on the free running LO.

The unlocked LO frequency $\tilde{\omega}_{\rm LO}(t)$ is affected by stochastic fluctuations that are characterized by a power spectral density $S(f) = h_{\rm LO}/f^{\alpha}$, with $\alpha=0$
for white noise (also often indicated as frequency noise) and 
$\alpha =1$ for flicker (or pink) noise, where $h_{\rm LO}$ is a prefactor. The accumulated phase during a time $T$ is $\tilde{\theta}(T) = \int_T dt ~  \delta\tilde{\omega}_{\rm LO}(t)$, where 
$\delta\tilde{\omega}_{\rm LO}(t) = \tilde{\omega}_{\rm LO}(t) - \omega_0$.
The quantity $\tilde{\theta}(T)$ is a stochastic variable with a Gaussian statistical distribution of zero mean, $\mathcal{E}_{\tilde{\omega}}[\tilde{\theta}(T)]=0$, and variance 
$v_\alpha(T)^2 = \mathcal{E}_{\tilde{\omega}}[\tilde{\theta}_{\rm LO}(T)^2]$, where ${\mathcal{E}}_{\tilde{\omega}}$ indicates statistical averaging over LO fluctuations.
For white noise, we have
$v_0(T)^2=\gamma_{\rm LO}T$, 
where the dephasing rate is related to $h_{\rm LO}$ as  $\gamma_{\rm LO}/\omega_0 =  h_{\rm LO} \omega_{\rm LO}/2$.
For flicker noise, we have 
$v_1(T)^2 = (\gamma_{\rm LO} T)^2$, 
where $\gamma_{\rm LO}/\omega_0 =   \sqrt{h_{\rm LO} 2 \chi \log 2}$ and $\chi=1.4$ is determined numerically (see Appendix for details on the numerical simulations).

To stabilize the LO frequency around $\omega_0$, one first uses the Ramsey interferometer to estimate the rotation angle that accumulates during the interrogation time $T$,
\be \label{deltaphi}
\theta(T) = \int_T dt ~ \delta \omega_{\rm LO}(t),
\ee
where $\delta \omega_{\rm LO}(t) = \omega_{\rm LO}(t) - \omega_0$.
From the estimated $\Theta(\mu)$, 
depending on the measurement result $\mu$
(for the joint Ramsey interferometer described in Sec. \ref{Sec.joint} we identify $\mu \equiv \{\mu_A, \mu_B \}$),
one obtains an estimate of the average LO frequency fluctuations,    
$\Theta(\mu)/T$~\cite{nota3}.
This value is subtracted from the signal $\delta\tilde{\omega}_{\rm LO}(t)$,
resulting in a feedback loop.
The estimation is repeated sequentially. 
In particular, during the $n$th Ramsey cycle, namely for times $(n-1)T \leq t \leq nT$
(with $n=2, ..., n_c$), the stabilized LO frequency is 
\be \label{correction}
\delta \omega_{\rm LO}(t) = \delta\tilde{\omega}_{\rm LO}(t) - \sum_{j=1}^{n-1} \frac{\Theta(\mu_j)}{T},
\ee  
where $\mu_j$ is the result of the $j$th measurement ($j=1, ..., n-1$).
Equation (\ref{correction}) provides the relation between the locked and the unlocked LO frequencies.
We point out that, in the joint Ramsey scheme of Fig.~\ref{Figure1} the two interferometers share the same LO
(and thus see the same LO fluctuations) and are characterized by the same atomic transition, 
interrogation time and number of particles, such that the phase shift $\theta(T)$ is common to both.

%If $T$ is sufficiently small, then $\theta(T)$ is very likely to be 
%within the inversion region $[-\ell, \ell]$ where an unbiased estimation is guaranteed, that is $\ell = \pi/2$ for the single-Ramsey interferometer and 
% $\ell = \pi$ for the joint-Ramsey interferometer, as discussed above.
%Increasing $T$, there is an increasing probability that the stochastic phase shift is outside the inversion region, namely $\vert \theta \vert > \ell$, resulting in an inaccurate correction of the LO. 
In the following, we evaluate the stability of the clock using the Allan variance, which is a common figure of merit.
In order to compare the schemes based on single and joint Ramsey interferometry, it is crucial to evaluate accurately the
optimal interrogation time.
In particular, we introduce an expression for the calculation of the average Allan variance that avoids numerical instabilities due to phase slips and allows to obtain analytical (for white LO noise) of semi-analytical (for correlated LO noise) results. 
The clock schemes based on single and joint Ramsey interferometry are discussed in Sec.~\ref{Sec.Single} and \ref{Sec.Joint}, respectively.
Notice that we neglect here dead times between Ramsey interrogations (this approximation is lifted in Sec.~\ref{Sec3_DT}).
We also assume that atomic decoherence occurs on time scales much longer than LO dephasing and therefore assume that LO decoherence 
is the only relevant noise source in the experiment.  
  
\subsection{Allan variance and phase slips}
\label{Sec.Allan}

{\it Allan variance.} 
We introduce the (dimensionless) fractional time-averaged frequency offset
\be \label{yAllan}
y_n(T, \mu_n) = \frac{\theta_n(T) - \Theta(\mu_n)}{\omega_0 T},
\ee
given by the difference between the accumulated rotation angle
$\theta_n(T) =  \int_{(n-1)T}^{nT} dt ~ \delta \omega_{\rm LO}(t)$ during the $n$th Ramsey cycle, 
and its estimate value $\Theta(\mu_n)$~\cite{nota3}. 
We quantify the stability of the clock by the fluctuations of the average $\tfrac{1}{n_c} \sum_{n=1}^{n_c} y_n(T, \mu_n)$ in $n_c$ Ramsey cycles.
In particular, the two-points variance (commonly indicated as Allan variance~\cite{Allan1966, Vanier1992, Wynands, Riehle2004}) 
\be \label{Allan0}
\sigma^2_{n_c} = \frac{1}{2n_c (n_c-1)} \sum_{n=1}^{n_c-1} (y_{n+1} - y_n)^2
\ee
is generally considered in order to estimate the stability due to stochastic noise since constant systematic errors cancels in Eq.~(\ref{Allan0}). 

In the absence of strong constant biases (such as those induced by phase slips), correlations between consecutive measurements of $y_n$ can be generally neglected:
$\sum_{n=1}^{n_c-1} y_{n+1} y_n \ll \sum_{n=1}^{n_c-1} y_n^2 \approx \sum_{n=1}^{n_c-1} y_{n+1}^2$. 
Indeed, although the frequency noise may be correlated, 
the phase estimations are uncorrelated, which makes 
negligible the correlations between $y_n$ and $y_{n+1}$ when compared to their fluctuations. 
We thus find 
\be \label{Allan1}
\sigma^2_{n_c} = \frac{1}{(n_c-1)n_c} \sum_{n=1}^{n_c-1} y_n^2.
\ee
For sufficiently large $n_c$, we can replace the weighted average by a statistical average -- that we indicate as ${\mathcal{E}}_{\mu, \theta}$ -- 
over both random frequency fluctuations 
(or, equivalently, random values of $\theta$) and random measurement results $\mu$:
\be \label{Allan2}
\sigma^2_{n_c} = \frac{1}{\omega_0^2 T^2 n_c} {\mathcal{E}}_{\mu, \theta}\Big[ \big( \theta(T) - \Theta(\mu) \big)^2 \Big].
\ee
Introducing the joint probability distribution $P_T(\mu, \theta)$, which depends, in general, on the Ramsey time $T$, we can write Eq.~(\ref{Allan2}) as
\be \label{Allan3}
\sigma^2_{n_c} = \frac{1}{\omega_0^2 T^2 n_c} \int {\rm d} \theta \sum_{\mu} P_T(\mu, \theta) \big( \theta - \Theta(\mu) \big)^2.
\ee 
Finally, using the basic conditional probability relation $P_T(\mu, \theta) = P(\mu \vert \theta) P_T(\theta)$ and
$\sum_{\mu} P(\mu \vert \theta) \big( \theta - \Theta(\mu) \big)^2 = \big(\Delta \Theta(\theta) \big)^2 + \big(\theta - \bar{\Theta}(\theta)\big)^2$, we arrive at the equation 
\be  \label{Allan4}
\sigma^2_{n_c} = \frac{c_{T}^2}{\omega_0^2 T^2 n_c}, 
\ee
where 
\be \label{alphaT}
c_{T}^2 = \int {\rm d} \theta\,P_T(\theta) \, \Big[ \big( \Delta \Theta(\theta) \big)^2  + \big(\theta - \bar{\Theta}(\theta) \big)^2 \Big]
\ee
is a weighted average of the mean square error. 
Equation (\ref{Allan4}) links $\sigma^2_{n_c}$, the estimator variance $\big( \Delta \Theta \big)^2$ and the bias $\vert \theta - \bar{\Theta}(\theta) \vert$.
In particular, $c_T^2$ fulfils the Cram\`er-Rao bound~\cite{CramerBOOK, Rao1945}:
\be \label{CRLB}
c_{T}^2 \geq \int {\rm d} \theta\,P_T(\theta) \, \bigg[ \frac{1}{F(\theta)} \bigg(\frac{d \bar{\Theta}}{d \theta} \bigg)^2  + \big(\theta - \bar{\Theta}(\theta) \big)^2 \bigg],
\ee
where $F(\theta) = \sum_{\mu} \tfrac{1}{P(\mu \vert \theta)} \Big(
\tfrac{P(\mu \vert \theta)}{d \theta} \Big)^2$ is the Fisher information.
The bound (\ref{CRLB}) holds for any estimator $\Theta(\mu)$.
For unbiased estimators, namely $\bar{\Theta}(\theta) = \theta$, the bound (\ref{CRLB}) equals the weighted average of the inverse Fisher information, 
$c_{T}^2 \geq \int d\theta \,P_T(\theta)/F(\theta)$.
By combining Eqs.~(\ref{Allan4}) and (\ref{CRLB}) we obtain a lower bound to the Allan variance, although its saturation is not guaranteed in general. The optimization of $F(\theta)$ over all possible positive operator-valued measure defines the quantum Fisher information, $F(\theta) \leq F_{\rm Q}(\theta)$~\cite{HelstromBOOK, BraunsteinPRL1994}, that depends only on the probe state and interferometer transformation.
We thus obtain $c_{T}^2 \geq \int d\theta \,P_T(\theta)/F_Q(\theta)$ for unbiased estimators although the saturation of the bound is not guaranteed since  the optimal measure for which the equality $F(\theta) = F_{\rm Q}(\theta)$ holds depend, in general, on $\theta$~\cite{PezzeREVIEW}.

{\it Phase slips.} Equation (\ref{Allan4}) gives the Allan variance for a sequence of $n_c$ Ramsey cycles in the absence of phase slips, 
namely if $\vert \theta_n \vert \leq \ell$ for $n=1, ..., n_c$, where $2\ell$ indicates the total width of the inversion region (e.g. $\ell=\pi/2$ for the single Ramsey interferometer, see Sec. \ref{Sec.SingleRamsey}). 
If a phase slip occurs at the $n$th Ramsey cycles, namely  $\vert \theta_n \vert > \ell$, 
the estimation method leads to a bias. It is generally unlikely that the frequency can be stabilized around $\omega_0$ if we interrogate further the LO.
Instead, additional phase slips may occur, which further increase the estimation bias and make an analytical calculation of the Allan variance cumbersome \cite{noteAllan}.  
In the following, to avoid numerical instabilities, the LO is stopped after the occurrence of a single phase slip.
In this case, the number or Ramsey cycles $n_c$ during which no phase slip occurs is thus a stochastic variable. 
Let us thus indicate as $P_T(n_c)$ the probability that a phase slip occurs at the $n_c$th Ramsey cycle.
This is given by 
\be \label{Pstop}
P_T(n_c) = 
\begin{cases}
1-  2 \int_{0}^{\ell} d\theta~P_T(\theta) & {\rm for} \quad n_c=1, \\
P_{n_c}(T)  \times \prod_{n=1}^{n_c-1} \Big(1-P_{n}(T) \Big) & {\rm for} \quad n_c>1,
\end{cases}
\ee
where $P_{n}(T) = 1-2\int_{0}^{\ell} d \theta_n P_T(\theta_n)$ is the probability that $\vert \theta_n(T) \vert > \ell$. 

In the case of a LO with white noise frequency fluctuations, the values of $\theta_n$ are uncorrelated:
the probability $P_{n}(T)$ is thus the same for all Ramsey cycles and it is given by $p(T) = 1-  2 \int_{0}^{l} d\theta~P_T(\theta)$.
In this case, Eq.~(\ref{Pstop}) becomes
\be \label{Pstop_wn}
P_{T}(n_c) = \big(1-p(T)\big)^{n_c-1} \times p(T).
\ee
While the above equations hold for any probability distribution $P_T(\theta)$, a relevant case is the Gaussian
\be \label{PTtheta}
P_T(\theta) = \frac{e^{-\tfrac{\theta^2}{2 v_T^2}}}{\sqrt{2 \pi v_T^2}}
\ee
where the variance $v_T^2$ depends on the interrogation time, 
as discussed above.
In this case we have 
\be \label{Pps}
p(T) =
1 - {\rm Erf}\Bigg(\frac{\ell}{\sqrt{2} v_T}\Bigg),
\ee
where ${\rm Erf}$ is the error function. 

{\it General expression for the Allan variance in presence of phase slips.}
Let us introduce the total averaging time $\tau$. 
The quantity $\floor{\tau/T}$, indicating the largest integer less than or equal to $\tau/T$, sets the maximum number of Ramsey cycles (neglecting dead times, see below).
Our general expression for the Allan variance is
\be \label{AllanGeneral}
\sigma^2 =  Q_{T}(\floor*{\tau/T}) \sigma^2_{\floor*{\tau/T}} + \sum_{n_c=2}^{\floor*{\tau/T}} P_{T}(n_c) \sigma^2_{n_c}.
\ee
The first term corresponds to the case where no phase slip happens in any of the
$\floor*{\tau/T}$ Ramsey cycles: this occurs with probability
$Q_{T}(\floor*{\tau/T}) = \prod_{n=1}^{\floor*{\tau/T}} \big(1 - P_n(T)\big)$.
The second term in Eq. (\ref{AllanGeneral}) corresponds to a statistical average 
of $\sigma^2_{n_c}$ with $P_{T}(n_c)$ being the probability that a phase slip at the 
$n_c$th Ramsey cycle, as given in Eq. (\ref{Pps}).
The sum starts from $n_c=2$ because at least 
two Ramsey cycles are required to calculate the Allan variance according to Eq.~(\ref{Allan0}).
Finally, using Eq.~(\ref{Allan4}), we find
\be \label{AllanFinal}
\sigma^2 = \frac{c^2_{T}}{\omega_0^2 T^2}
 \Bigg( \frac{Q_{T}(\floor*{\tau/T})}{\floor*{\tau/T}} + \sum_{n_c=2}^{\floor*{\tau/T}} \frac{P_{T}(n_c)}{n_c} \Bigg).
\ee
This equation
is characterized by two limits. 
If the interrogation time $T$ and the ratio $\tau/T$ are sufficiently short such that $P_{T}(n_c) \approx 0$ for $n_c = 1, ..., \floor*{\tau/T}$ and thus $Q_{T}(\tau/T) \approx 1$, then the effect of phase slips is negligible and we obtain
\be \label{shortAV}
\sigma^2 = \frac{c_T^2}{\omega_0^2 T \tau},
\ee
where we have approximated $T \times \floor*{\tau/T}\approx \tau$.
This equation recovers the characteristic scaling $\sigma^2 \sim 1/(T\tau)$.
In the opposite regime, when the term $Q_{T}(\floor*{\tau/T})/(\floor*{\tau/T})$ is negligible (e.g. in the limit $\tau/T \to \infty$), we obtain 
\be \label{longAV}
\sigma^2 = \frac{c_T^2}{\omega_0^2 T^2}  \sum_{n_c=2}^{\floor*{\tau/T}} \frac{P_{T}(n_c)}{n_c}.
\ee
In this case, phase slips dominate the calculation of the Allan variance and we find that $\sigma^2$ does not scale with $\tau$: in this case, changing the total averaging time has no effect on the Allan variance since there is a large probability that a phase slip occurs for $n_c \leq \floor*{\tau/T}$.

%%%%%%%%%%%%%%%%%%%%%%%%%%%%%%%%%%%%%%%%%%%%%%%
%% Figure 3
%%%%%%%%%%%%%%%%%%%%%%%%%%%%%%%%%%%%%%%%%%%%%%%
\begin{figure}[t!]
\includegraphics[width=8.5cm]{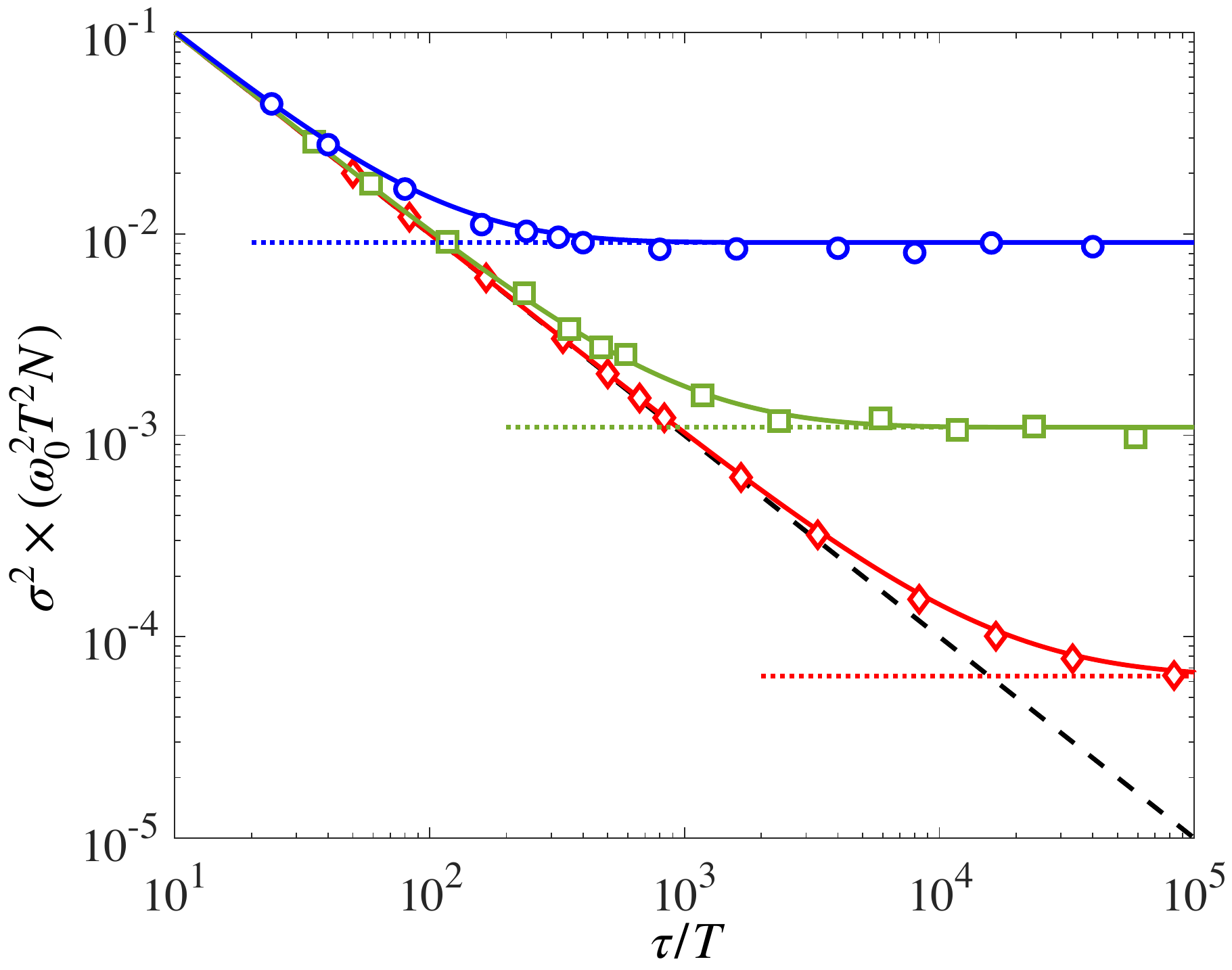}
\caption{
Allan variance as a function of $\tau/T$ for white LO noise and a clock based on a single Ramsey interferometer. 
Symbols are results of ab initio numerical simulation for different values of the Ramsey time:
$\gamma_{0}T = 0.25$ (blue circles), 
$\gamma_{0}T = 0.17$ (green squares)
and $\gamma_{0}T = 0.12$ (red diamonds).
Solid lines are Eq.~(\ref{sigma_wn}), the dashed line is the SQL, Eq.~(\ref{sigma_smallT}), while the 
dotted lines are the asymptotic $\tau/T \to + \infty$ prediction of Eq.~(\ref{sigma_largeT_wn}).
Here, $N=1000$.}
\label{Figure3}
\end{figure}
%%%%%%%%%%%%%%%%%%%%%%%%%%%%%%%%%%%%%%%%%%%%%%%
%%%%%%%%%%%%%%%%%%%%%%%%%%%%%%%%%%%%%%%%%%%%%%%
%%%%%%%%%%%%%%%%%%%%%%%%%%%%%%%%%%%%%%%%%%%%%%%

%%%%%%%%%%%%%%%%%%%%%%%%%%%%%%%%%%%%%%%%%%%%%%%
%% Figure 4
%%%%%%%%%%%%%%%%%%%%%%%%%%%%%%%%%%%%%%%%%%%%%%%
\begin{figure*}[t!!]
\includegraphics[width=\textwidth]{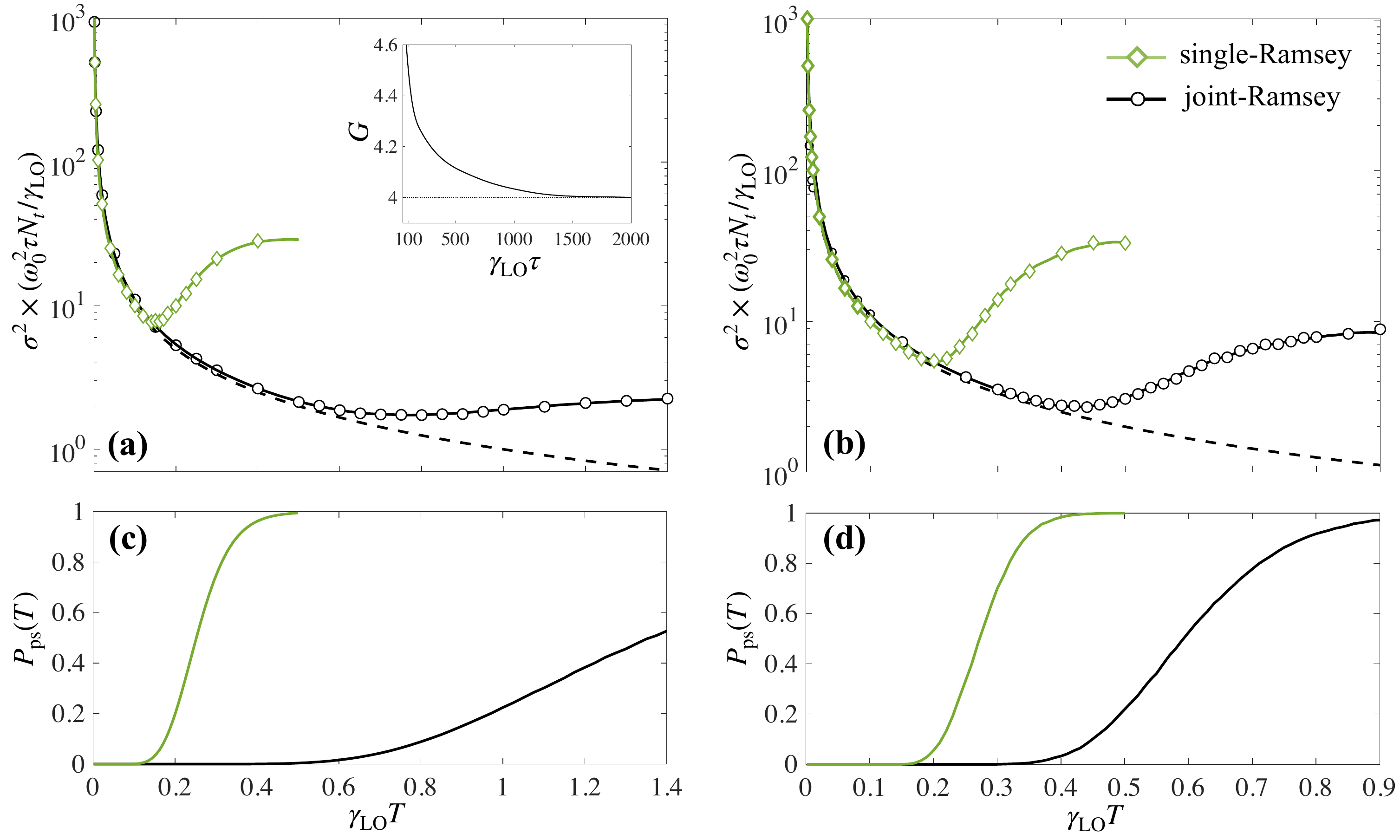}
\caption{Allan variance (multiplied by $\omega_0^2 \tau N / \gamma_{\rm LO}$) 
as a function of the Ramsey time $\gamma_{\rm LO} T$ for white (a) and flicker (b) LO noise, respectively.
Symbols are results of ab-initio numerical simulations: 
green diamonds are obtained for a single-Ramsey clock with $N=2000$ particles, while 
black circles are obtained for a joint-Ramsey clock with $1000$ particles in each interferometer ($N_t = 2000$ is the total number of particles used in both clocks).
The dashed line in both panels is the SQL, $\sigma^2 = 1/(\omega_0^2 T \tau N_t)$.
The solid lines in both panels are expected results. 
For the white noise case, the solid green line is Eq.~(\ref{sigma_wn}), while the solid black line is Eq. (\ref{sigma_wn_joint}).
For the flicker noise, the solid lines are 
semi-analytical predictions given by Eq. (\ref{AllanFinal}) where the quantities $Q(\tau/T)$, $P_T(n_c)$ and $c_T^2$ are obtained numerically and independently 
from the calculation of the Allan variance.
The inset of panel (a) shows the parameter $G$ defined in Eq. (\ref{Rsigma}) for the white noise LO case, as a function of the total averaging time $\gamma_{\rm LO} \tau$.
The solid line is obtained analytically and shows that the gain in the noiseless case converges to $G=4$ for large $\tau$. 
Panels (c) and (d) show the probability of phase slip, Eq.~(\ref{Ppstot}), 
as a function of $\gamma_{\rm LO} T$, for white and flicker noise, respectively.
The solid lines in panel (c) is Eq.~(\ref{Ppstotwn}).
In all panels, $\gamma_{\rm LO} \tau = 100$.
}
\label{Figure4}
\end{figure*}
%%%%%%%%%%%%%%%%%%%%%%%%%%%%%%%%%%%%%%%%%%%%%%%
%%%%%%%%%%%%%%%%%%%%%%%%%%%%%%%%%%%%%%%%%%%%%%%
%%%%%%%%%%%%%%%%%%%%%%%%%%%%%%%%%%%%%%%%%%%%%%%

\subsection{Allan variance and phase slips for the single-Ramsey clock protocol}
\label{Sec.Single}

In Fig.~\ref{Figure3} we plot the Allan variance for the clock protocol based on a single Ramsey interferometer, as a function of $\tau/T$. 
Different symbols are results of ab initio numerical simulations for different fixed values of Ramsey time $T$. 
For values of $\theta$ inside the inversion region $[-\ell, \ell]$ ($\ell=\pi/2$ in this case), we have $\vert \theta - \bar{\Theta}_A \vert \ll \Delta \Theta_A(\theta) = 1/\sqrt{N}$, see 
Fig.~\ref{Figure2}(a), and thus $c_T^2 = 1/N$. 
In particular, according to 
Eq. (\ref{shortAV}), for sufficiently short values of $\tau/T$, we recover
the standard quantum limit (SQL) ~\cite{ItanoPRA1993, Salomon1999, LudlowRMP2015}
\be \label{sigma_smallT}
\sigma^2 = \frac{1}{\omega_0^2 N T \tau}.
\ee
For large values of $\tau/T$, the Allan variance is characterized by a saturation, as predicted by Eq. (\ref{longAV}), 
\be \label{sigma_largeT}
\sigma^2 = \frac{1}{\omega_0^2 N T^2} \sum_{n_c=2}^{\floor*{\tau/T}} \frac{P_{T}(n_c)}{n_c}.
\ee
For white noise, Eq.~(\ref{AllanFinal}) can be calculated using Eq.~(\ref{Pstop_wn}) and $c_T^2 = 1/N$. This gives
\be \label{sigma_wn}
\sigma^2 = \frac{1}{\omega_0^2 T^2 N} \Bigg[ \frac{p(T)}{1- p(T)} \sum_{n_c=2}^{\floor*{\tau/T}} \frac{\big(1-p(T)\big)^{n_c}}{n_c} + \frac{\big(1-p(T)\big)^{\floor*{\tau/T}}}{\tau/T} \Bigg],
\ee
where we have used 
$Q_{T}(\floor*{\tau/T}) = \big(1-p(T)\big)^{\floor*{\tau/T}}$
and $p(T)$ is calculated according to Eq. (\ref{Pps}). In the limit $\tau/T \to +\infty$, Eq.  (\ref{sigma_wn}) becomes
\be \label{sigma_largeT_wn}
\sigma^2 = \frac{1}{\omega_0^2 T^2 N} \Bigg[ \frac{p(T)}{1- p(T)} \log \frac{1}{p(T)} - p(T) \Bigg].
\ee
As shown in Fig. \ref{Figure3}, the numerical results agree very well with the analytical prediction [solid lines, given by Eq. (\ref{sigma_wn})]. The dotted lines is the asymptotic plateau
Eq.~(\ref{sigma_largeT_wn}).

To study the Allan variance as a function of the Ramsey time $T$, 
we fix the total averaging time $\tau$.
In this case, by increasing $T$, we find a transition between Eq.~(\ref{sigma_smallT}) and Eq.~(\ref{sigma_largeT}):
while Eq.~(\ref{sigma_smallT}) predicts that the Allan variance decreases as $\sigma^2 \sim 1/T$,
Eq.~(\ref{sigma_largeT}) increases as a function of $T$.
The bending knee of the Allan variance 
(that identifies the absolute stability of the clock) is clearly shown in Fig.~\ref{Figure4}(a) and (b), obtained for white and flicker LO noise, respectively. 
Green diamonds are results of ab-initio numerical simulations. 
The solid line in panel (a) is the analytical prediction Eq.~(\ref{sigma_wn}) for white noise, while for 
flicker noise (b) the solid green line is obtained by calculating $P_{T} (n_c)$ and $Q_{T}(\tau/T)$ numerically (and independently from the numerical simulations of $\sigma^2$).

In Fig.~\ref{Figure4} (c) and (d) we show the overall probability that a phase slip occurs in one of the $\floor*{\tau/T}$ Ramsey cycles, namely
\be \label{Ppstot}
P_{{\rm ps}}\big(T,\floor*{\tau/T}\big) = \sum_{n_c=1}^{\floor*{\tau/T}} P_{T}(n_c) = 1-Q_{T}(\floor*{\tau/T}),
\ee
as a function of the Ramsey time. 
Panel (c) in Fig.~\ref{Figure4} is obtained for white noise, while panel (d) for flicker noise. 
In the case of white noise, Eq.~(\ref{Ppstot}) can be evaluated analytically as 
\be \label{Ppstotwn}
P_{{\rm ps}}\big(T,\floor*{\tau/T}\big) = 1 - \big(1-p(T) \big)^{\floor*{\tau/T}},
\ee
given by the solid line in Fig. \ref{Figure4}(c). 
We observe that the bending of the Allan variance is obtained 
when $P_{{\rm ps}}(T,\floor*{\tau/T}) \approx 0.05$ (and similarly for flicker noise). 
It should be noticed that the optimal Ramsey time corresponding to the minimum of the Allan variance has a 
slight dependence on the total interrogation time $\tau$ (in particular it decreases with $\tau$).
This is due to the fact that increasing $\tau$, for a fixed Ramsey time $T$, the number of cycles increases and thus 
the probability to have a phase slip increases as well, as shown in Eq.~(\ref{Ppstotwn}).
In particular, in the limit 
$\tau/T \to \infty$, we have $P_{\rm ps}(T, \floor*{\tau/T}) \to 1$ whenever $p(T)>0$.

\subsection{Allan variance and phase slips for the joint-Ramsey clock protocol}
\label{Sec.Joint}

The Allan variance for the joint-Ramsey clock can be obtained following the discussion of Sec.~\ref{Sec.Allan}.
To be explicit, our semi-analytical prediction is 
given by Eq.~(\ref{AllanFinal})
with $c_T^2 = \int {\rm d} \theta\,P_T(\theta) \, \big[ \big( \Delta \Theta(\theta) \big)_{AB}^2  + \big(\theta - \bar{\Theta}_{AB}(\theta) \big)^2 \big]$.
The major difference with respect to the  single-Ramsey clock is the size of the inversion region $[-\ell, \ell]$ that is identified here with $\ell = \pi - 4/\sqrt{N}$ (rather than $\ell = \pi/2$ as in the case of a single Ramsey interferometer). 
The factor $4/\sqrt{N}$ is arbitrary and guarantees a small probability of biased estimation close to the $\theta = \pm \pi$, see Fig. \ref{Figure2}(b).
In particular, for the white noise case, we have 
\be \label{sigma_wn_joint}
\sigma^2 = \frac{c_T^2}{\omega_0^2 T^2} \Bigg[ \frac{p(T)}{1- p(T)} \sum_{n_c=2}^{\floor*{\tau/T}} \frac{\big(1-p(T)\big)^{n_c}}{n_c} + \frac{\big(1-p(T)\big)^{\floor*{\tau/T}}}{\tau/T} \Bigg],
\ee
with $p(T)$ given in Eq. (\ref{Pps}) with $\ell = \pi - 4/\sqrt{N}$.

In Fig.~\ref{Figure4}(a) and (b) we show the results of numerical calculations of the Allan variance for the joint-Ramsey clock, in the case of white and flicker LO noise, respectively.
In the figure, the single-Ramsey clock (green diamonds) and the joint-Ramsey scheme (black circles) are compared for the same total number of particles $N_t$
(in the joint scheme the total number of particles is $N_t=2N$).
Notice that the black circles stay slightly above the dashed line, giving the SQL $\sigma^2_{\rm SQL} = 1/(\omega_0^2 T \tau N_t)$.
This effect is due to the 
mean square error $\big(\Delta  \Theta_{AB}\big)^2 + \big(\theta - \bar{\Theta}_{AB}(\theta) \big)^2$, which is relevant to calculate the quantity $c_T^2$ in Eq.~(\ref{AllanFinal}), being slightly above $1/N_t$ for $\theta$ close to $0$ and $\pm \pi/2$, see Fig. \ref{Figure2}(b). 
This is a minor effect that can nevertheless be seen in the numerical simulations. For both white and flicker noise, the numerical results are well reproduced by semi-analytical findings (solid lines) obtained using Eq.~(\ref{AllanFinal}).

As shown in Fig. \ref{Figure4}(a) and (b), using the joint-Ramsey interrogation, the minimum of the Allan variance is reached for longer interrogation times, with respect to the single-Ramsey clock. 
This corresponds to an increase of the absolute stability that can be quantified by the gain factor 
\be \label{Rsigma}
G = \frac{(\min_{T} \sigma^2)_{\rm single}}{(\min_{T} \sigma^2)_{\rm joint}},
\ee
given by the ratio between the Allan variance for the single-Ramsey clock (with $N_t$ particles) and that of the joint protocol (with $N$ particles in each state, $N_t=2N$ in total), each optimized with respect to the Ramsey time.
For while LO noise we obtain that 
$G\approx 4$. In particular, the inset of Fig. \ref{Figure4}(a) shows $G$ as a function of $\gamma_{\rm LO} \tau$.
For $\gamma_{\rm LO} \tau  = 100$ corresponding to the results shown in the main panel of Fig. \ref{Figure4}(a), we obtain $G \approx 4.5$. 
For large values of $\tau$ we obtain that $G$ converges to the value 4. 
For flicker noise, we obtain $G\approx 2$ with a weak dependence on $\tau$ (not shown).
The increase of stability obtained with the joint clock is directly related to the smaller probability of phase slips, as shown in Fig. \ref{Figure4}(c) and (d).

%%%%%%%%%%%%%%%%%%%%%%%%%%%%%%%%%%%%%%%%%%%%%%%
%% Figure 5
%%%%%%%%%%%%%%%%%%%%%%%%%%%%%%%%%%%%%%%%%%%%%%%
\begin{figure}[t!!]
\includegraphics[width=\columnwidth]{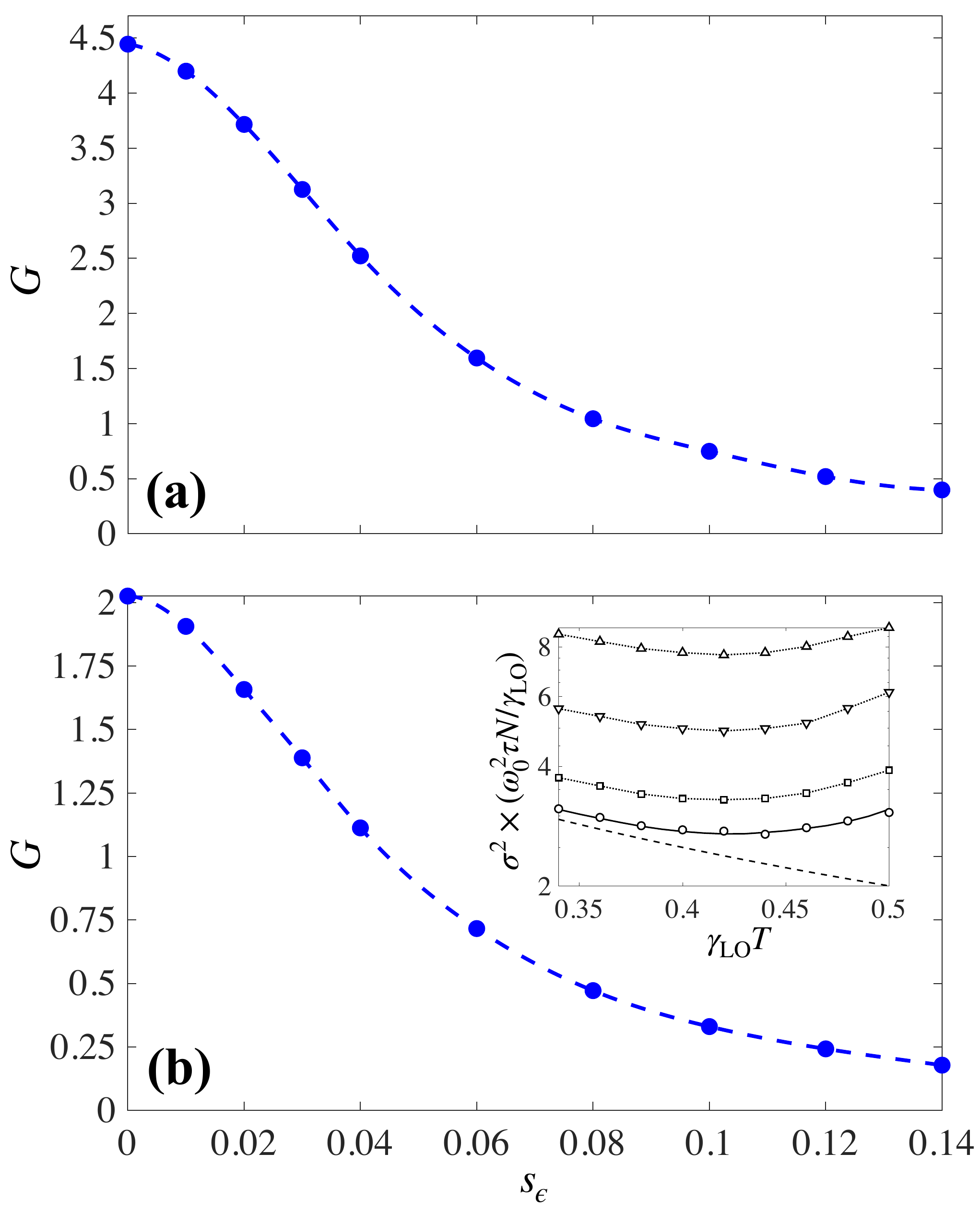}
\caption{Absolute stability gain $G$, Eq.~(\ref{Rsigma}), 
as a function of $s_{\epsilon}$ quantifying the imperfect alignment of the two probe states of the joint-Ramsey clock.
Panel (a) is obtained for white LO noise, while panel (b) for flicker noise. 
Blue circles are results of numerical simulations, the dashed line is a guide to the eye.
The inset of panel (b) shows the Allan variance as a function of the Ramsey time $\gamma_{\rm LO}T$.
Different symbols are obtained for different values of $\epsilon$:
$s_{\epsilon}=0$ (circles),  
$s_{\epsilon}=0.02$ (squares),
$s_{\epsilon}=0.04$ (downward-pointing triangle), and 
$s_{\epsilon}=0.06$ (upward-pointing triangle).
The solid line is Eq.~(\ref{AllanFinal}),
the dotted lines are a guide to the eye and  
the dashed line is the SQL.
In all panels, $\gamma_{\rm LO}\tau = 100$, and the total number of particles is $N_t=2000$.
}
\label{Figure5}
\end{figure}
%%%%%%%%%%%%%%%%%%%%%%%%%%%%%%%%%%%%%%%%%%%%%%%
%%%%%%%%%%%%%%%%%%%%%%%%%%%%%%%%%%%%%%%%%%%%%%%
%%%%%%%%%%%%%%%%%%%%%%%%%%%%%%%%%%%%%%%%%%%%%%%  

%%%%%%%%%%%%%%%%%%%%%%%%%%%%%%%%%%%%%%%%%%%%%%%
%% Figure 6
%%%%%%%%%%%%%%%%%%%%%%%%%%%%%%%%%%%%%%%%%%%%%%%
\begin{figure*}[t!!]
\includegraphics[width=2\columnwidth]{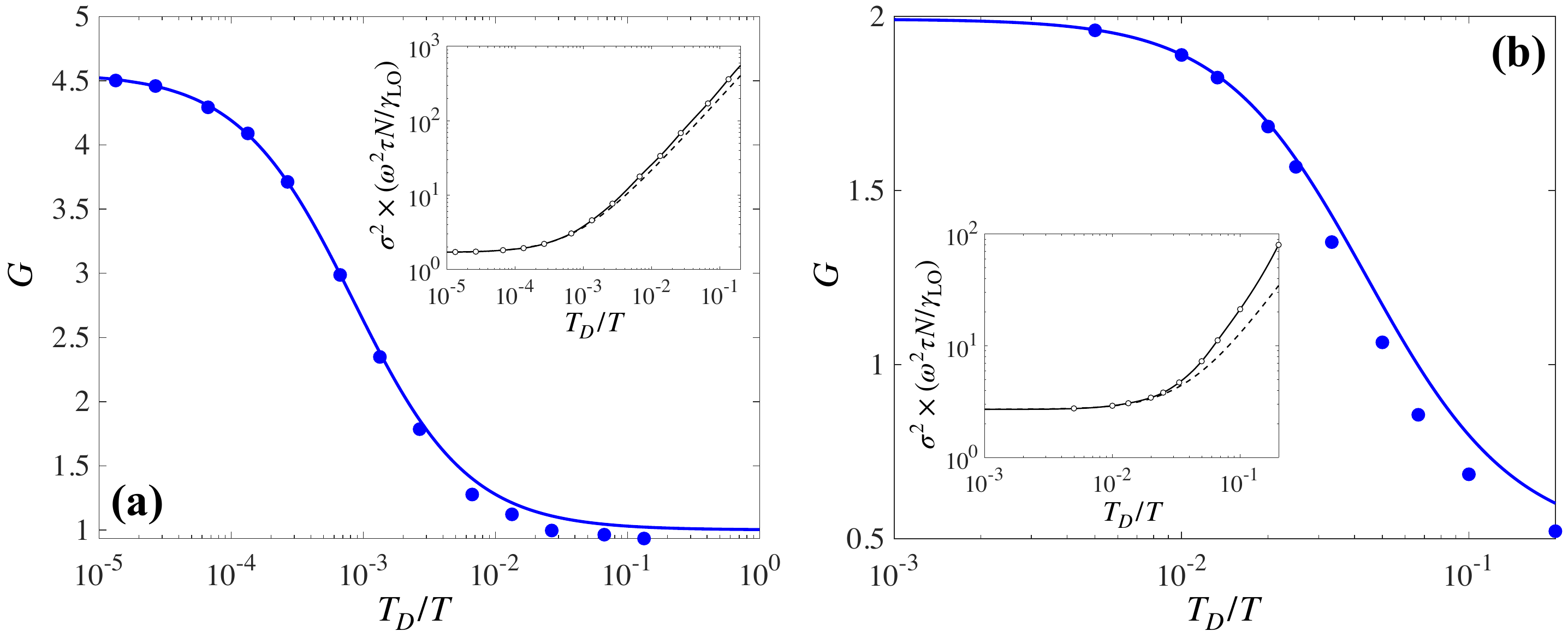}
\caption{Stability gain factor Eq.~(\ref{Rsigma}) as a function of  
the dead time $T_D/T$ for white (a) and flicker (b) noise, respectively.
Here the Ramsey time $T$ is set to that minimizing the Allan variance for $T_D=0$, see Fig.~\ref{Figure4} and is different for the single- and the joint-Ramsey scheme.
Circles are results of numerical simulations, while the solid line is the analytical prediction of Eqs.~(\ref{Dick}) and (\ref{AllanD}).
The inset in both panels shows the Allan variance $\sigma^2$ as a function of $T_D/T$ for the joint-Ramsey clock where $T$ is the Ramsey time minimizing the Allan variance for $T_D=0$.
Circles are results of numerical simulations. 
The dashed line is Eq.~(\ref{AllanD}) with $\sigma_D^2$ given by Eq.~(\ref{AllanWN})
for white LO noise [inset of panel (a)]
and by Eq.~(\ref{AllanPN})
for flicker LO noise [inset of panel (b)].
Here the total number of particles is $N_t=2000$, the averaging time is $\tau/T_C=100$.
}
\label{Figure6}
\end{figure*}
%%%%%%%%%%%%%%%%%%%%%%%%%%%%%%%%%%%%%%%%%%%%%%%
%%%%%%%%%%%%%%%%%%%%%%%%%%%%%%%%%%%%%%%%%%%%%%%
%%%%%%%%%%%%%%%%%%%%%%%%%%%%%%%%%%%%%%%%%%%%%%%

\section{Impact of possible imperfections}
\label{Sec4}

In this section we consider the impact of possible experimental imperfections in the implementation of the joint-Ramsey scheme.
Specifically, we release the assumption of perfect $\pi/2$ shift between the quantum states in the two interferometers and also 
consider the impact of number of particles fluctuations
and dead time.

\subsection{Imperfect alignment of the probe state}

We study here the joint Ramsey scheme of Fig. \ref{Figure1}(a) where the probe state of the Ramsey B interferometer is given by the statistical mixture
\be \label{rhoB}
\hat{\rho}_B = \int d\epsilon\, P(\epsilon)\,
\ket{\psi_{B}}_{\epsilon}\bra{\psi_{B}},
\ee
where 
$\ket{\psi_{B}}_\epsilon = (e^{-i (\pi/2-\epsilon)/2} \ket{\uparrow} + e^{i (\pi/2-\epsilon)/2} \ket{\downarrow})^{\otimes N}/2^{N/2}$,
and 
$P(\epsilon) = e^{-\epsilon^2/(2s^2_{\epsilon})}/\sqrt{2\pi s^2_{\epsilon}}$.
The state (\ref{rhoB}) reduces to Eq.~(\ref{CSS2}) for $s_\epsilon = 0$.
Here, the quantity $\epsilon$ is a stochastic value that accounts for the error in the precise alignment of the probe state in Ramsey B.
Figure \ref{Figure5}(a) and (b) show the gain Eq. (\ref{Rsigma}) as a function of $s_{\epsilon}$ for white and flicker LO noise, respectively.
As shown in the inset of Fig. \ref{Figure5}(b), the main effect of the imperfection in the state preparation is to increase the Allan variance, while the optimal Ramsey time remains approximately constant. 
The gain factor $G$ decreases while increasing $s_{\epsilon}$.
The condition to have 
values $G>1$ is more stringent in the case of flicker LO noise.

\subsection{Fluctuating number of particles}

We can study the case of number of particles fluctuations by replacing the coherent spin states (\ref{CSS1}) and (\ref{CSS2}) by 
a statistical distribution of coherent spin states of $N$ particles where $N$ is stochastic variable with 
mean $\bar{N}$ and fluctuations $\Delta N$.
The effect of number of particles fluctuations enters in our calculations via the requirement that the accumulated phase for the joint interferometer is sufficiently far from 
$\theta \lesssim \vert \pi - 4/\sqrt{N} \vert$ when interrogating a state of $N$ particles. 
In case of number of particles fluctuations, this conditions can be replaced by 
$\theta \lesssim \vert \pi - 4/\sqrt{\bar{N} + 4 \Delta N} \vert$ where the factor 4 is arbitrary and introduced so to avoid estimation biases for $\theta \approx \pm \pi$ (see discussion above). 
This introduces a very mild dependence of our results on $\Delta N$.
We thus expect that the joint interrogation method is robust against fluctuations of the number of atoms. 

\subsection{Dead times}
\label{Sec3_DT}

In common experimental realizations of atomic clocks, the interrogation of the atoms during a Ramsey time $T$ is followed by a dead time $T_D$ required for experimental
operations such as detection, loading, laser cooling, and state preparation, during which the LO is not interrogated by the atoms. 
Nevertheless, the dead time can be completely eliminated by synchronous interrogation of two atomic ensembles~\cite{BizeIEEE2000, TakamotoNATPHOT2011, BiedermannPRL2013, MeunierPRA2014, SchioppoNATPHOT2017}, 
in which the frequency correction derived from the interrogation of one system is applied during the interrogation of the second one.

There are two main consequences associated to the dead time:
1) The number of Ramsey cycles, for given Ramsey time $T$ and given total interrogation time $\tau$,
is now given by $\tau/T_C$ and decreases as $T_D$ increases, where $T_C = T+T_D$.
The main effect associated to the decrease of the number of Ramsey cycles is to replace
$\floor*{\tau/T}$ in Eq.~(\ref{AllanGeneral}) by $\floor*{\tau/T_C}$. 
For instance, the SQL, Eq.~(\ref{sigma_smallT}), is now replaced by \cite{LudlowRMP2015, ItanoPRA1993}
\be
\sigma^2_{\rm SQL} = \frac{1}{\omega_0^2 T^2 N} \frac{T_C}{\tau}.
\ee
The smaller number of cycles is also associated to a slight decrease of probability of phase slips, as discussed above.
This is however a minor effect.   
2) The other, more subtle, phenomenon associated to the dead time is the so-called Dick effect, 
as first discussed in Ref.~\cite{Dick1987}. 
In the presence of a dead time, Eq. (\ref{yAllan}) is replaced by
\be
y_n(T, \mu_n) = \frac{\theta_n(T) + \tilde{\theta}_n(T_D) - \Theta(\mu_n)}{\omega_0 T},
\ee
where $\theta_n(T) = \int_{(n-1)T_C}^{(n-1)T_C+T} d t~\delta\omega_{\rm LO}(t)$, 
is the phase accumulated during the 
interrogation time $T$
at the $n$th Ramsey cycle, and $\tilde{\theta}_n(T_D) = \int_{(n-1)T_C+T}^{nT_C} d t~\delta\omega_{\rm LO}(t)$
is the phase that accumulate during the dead time. 
The quantity $\Theta(\mu_n)$ is an estimate of $\theta_n(T)$ only.
In other words, the average phase $\theta_n(T_D)$ that accumulates during the dead time $T_C$ is not estimated by the inteferometer.
Effectively, the dead time results in the lack of information about the frequency spectrum of the LO due to the sampling process~\cite{Dick1987, AudoinIEE1998, SantarelliIEEE1998, QuessadaJOB2003, WestergaardIEEE2010}.
According to Refs. \cite{Dick1987, AudoinIEE1998}, the Dick effect alone is associated to an Allan variance 
\be \label{Dick}
\sigma^2_{D} = \frac{1}{\tau}  \sum_{k=1}^{+\infty} S(k/T_C) \bigg( \frac{\sin(k\pi d)}{k \pi d} \bigg)^2,
\ee
depending on the power spectral density of the free running LO taken at Fourier frequencies $k/T_C$, where $d = T/T_C$.
More explicitly, for white LO noise, Eq. (\ref{Dick}) becomes
\be \label{AllanWN}
\sigma^2_{D} = \frac{\gamma_{\rm LO}}{\omega_0^2 \tau} \frac{2}{ d^2} 
\sum_{k=1}^{+\infty} \frac{\sin^2(k\pi d)}{\pi^2 k^2},
\ee
while for flicker noise, we have
\be \label{AllanPN}
\sigma^2_{D} = \frac{\gamma_{\rm LO}}{\omega_0^2 \tau} \frac{\gamma_{\rm LO}}{2 \chi \log 2} \frac{T}{d^2} 
\sum_{k=1}^{+\infty} \frac{\sin^2(k\pi d)}{\pi^2 k^3}.
\ee
Following Ref.~\cite{SchulteNATCOMM2020},
we assume that the Allan variance associated to the Dick effect adds to the Allan variance due to the atomic interrogation. 
More explicitly, Eq.~(\ref{AllanFinal}), is now replaced by the sum of the two contributions:
\be  \label{AllanD}
\sigma^2 = \frac{c^2_{T}}{\omega_0^2 T^2}
 \Bigg( \frac{Q_{T}(\floor*{\tau/T_C})}{\floor*{\tau/T_C}} + \sum_{n_c=2}^{\floor*{\tau/T_C}} \frac{P_{T}(n_c)}{n_c} \Bigg) +  \sigma_{D}^2,
\ee
which, as discussed above, can be adapted for the single- or the joint-Ramsey interrogation.

In Fig.~\ref{Figure6} we show the results of numerical simulations of the different clock protocols in the presence of a dead time $T_D$.
The insets show the Allan variance of the joint-Ramsey clock as a function of $T_D/T$. Here $T$ is the optimal Ramsey time  minimizing the Allan variance in the case $T_D=0$, as shown in Fig.~\ref{Figure4}.
Circles are results of numerical simulations, the solid line is a guide to the eye, while the dashed line is the analytical prediction of Eqs.~(\ref{AllanD}) and (\ref{Dick}).
We notice a slight discrepancy between the analytical predictions and the numerical results when increasing $T_D/T$. This might be due to the relatively short $\tau$ and/or the assumptions leading to Eqs.~(\ref{AllanD}) and (\ref{Dick}). 
The main panels show the gain factor Eq.~(\ref{Rsigma}) as a function of $T/T_D$, where the Ramsey time $T$ is different for the single- and joint-Ramsey clocks.
Dots are numerical results, while the solid line is the prediction of Eqs.~(\ref{AllanD}) and (\ref{Dick}).
The Allan variance rapidly increases with $T_D$. As a direct consequence, the gain  factor decreases to values $G<1$ for $T_D/T \sim 10^{-2}$ for white LO noise [panel (a)] and $T_D/T \sim 5 \times 10^{-2}$ for flicker noise [panel (b)].

\section{Joint-Ramsey interrogation combined with spin-squeezing} 
\label{Sec5}

\begin{figure}[t!!]
\includegraphics[width=\columnwidth]{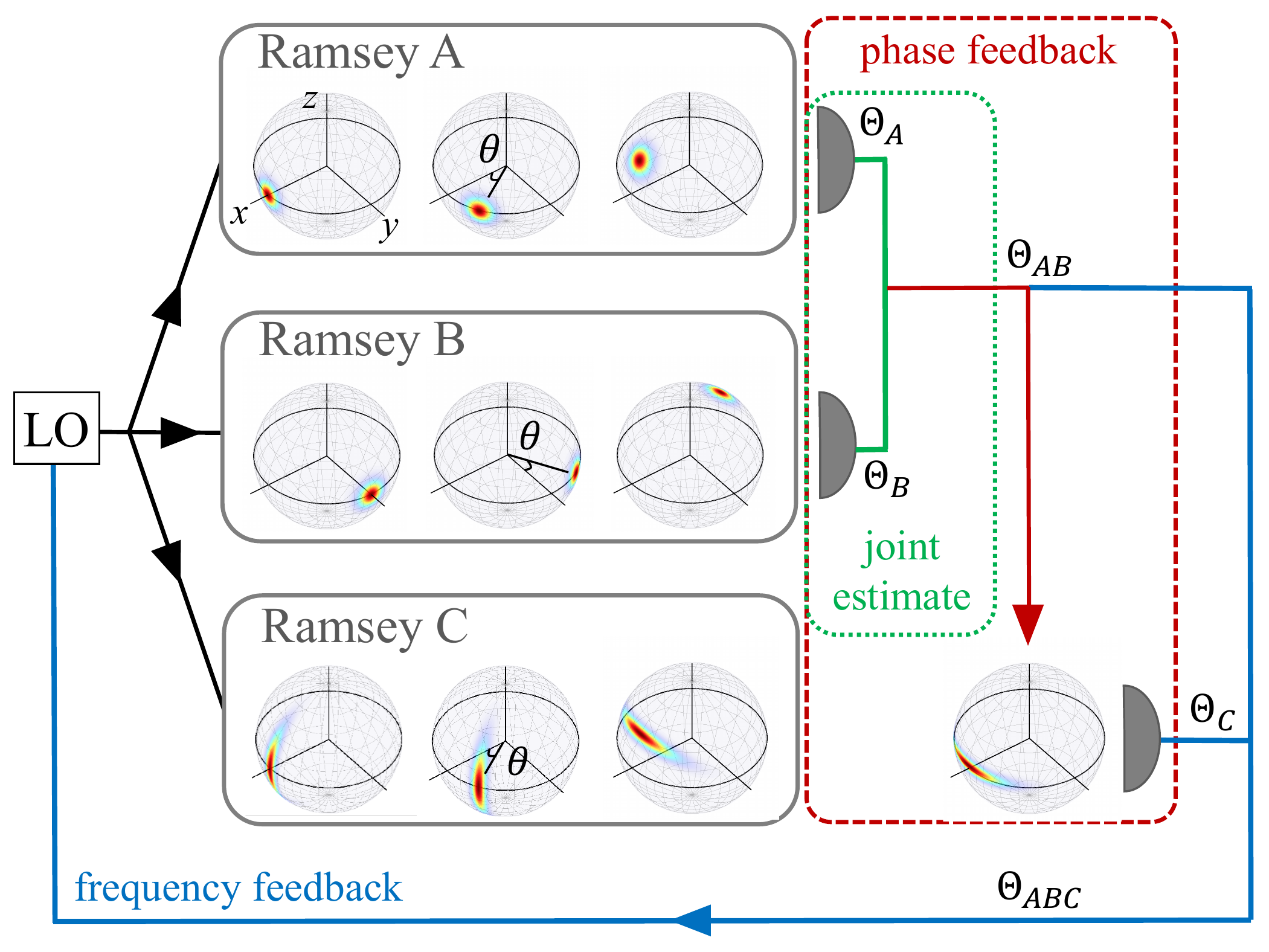}
\caption{Clock scheme combining joint-Ramsey interrogation and spin squeezing.
It consists of three Ramsey interferometers. 
The first two interferometers (Ramsey A and Ramsey B) use the coherent spin states (\ref{CSS1}) and (\ref{CSS2}), respectively.
The two estimates $\Theta_A$ and $\Theta_B$ are combined according to the discussion in Sec.~\ref{Sec.Joint}
(green dotted square and lines).
The joint estimate  $\Theta_{AB}$, Eq. (\ref{thest12}) is used, via a phase feedback (red dashed square and lines), to bring the spin-squeezed state toward its optimal working point, 
on the equator of the Bloch sphere.
The sum of $\Theta_{AB}$ and $\Theta_{C}$ giving $\Theta_{ABC}$, Eq. (\ref{thetaest123}), is used to steer the LO frequency via a frequency feedback (blue line).}
\label{Figure7}
\end{figure}

In the following, we combine the joint interrogation method discussed above with the approach proposed in Ref.~\cite{PezzePRL2020}
(see also \cite{PezzeARXIV}).
The overall clock scheme is shown in Fig.~\ref{Figure7}.
It consists of three Ramsey interferometers 
operating in parallel with 
the LO now interrogating three atomic states. 
We assume that the three states have the same number of particles $N$:
this is relevant for atomic clocks where, in order to increase the stability, 
one wants to use atomic ensembles that have maximum possible number of atoms, 
eventually limited by spatial constraints or the onset of decoherence affects associated to the large density.
The input state of Ramsey A is given by Eq.~(\ref{CSS1}), the input of Ramsey B is given by Eq.~(\ref{CSS2}), 
while the input of Ramsey C is 
\be \label{SSS3}
\ket{\psi_C} = \mathcal{N} \sum_{\mu = -N/2}^{N/2} e^{-\mu^2/(s^2N)} \ket{\mu}_y, 
\ee
where $\ket{\mu}_y$ are the eigenstate of $\hat{J}_y$ with eigenvalues $\mu=-N/2, -N/2+1, ..., N/2$, 
$\mathcal{N}$ provides the normalization and the parameter $s$ sets the variance $(\Delta \hat{J}_y)^2 = s^2 N/4$.
For $s<1$ the state (\ref{SSS3}) is spin squeezed~\cite{WinelandPRA1994, Louchet-ChauvetNJP2010, PezzeRMP2018, KrusePRL2016, PedrozoNATURE2020, HostenNATURE2016}, with metrological squeezing coefficient
$\xi^2 = N (\Delta \hat{J}_y)^2 / \mean{\hat{J}_x} \approx s^2 e^{1/(s^2 N)} < 1$~\cite{PezzePRL2020}.

%%%%%%%%%%%%%%%%%%%%%%%%%%%%%%%%%%%%%%%%%%%%%%%
%% Figure 7
%%%%%%%%%%%%%%%%%%%%%%%%%%%%%%%%%%%%%%%%%%%%%%%
%%%%%%%%%%%%%%%%%%%%%%%%%%%%%%%%%%%%%%%%%%%%%%%
%%%%%%%%%%%%%%%%%%%%%%%%%%%%%%%%%%%%%%%%%%%%%%%
%%%%%%%%%%%%%%%%%%%%%%%%%%%%%%%%%%%%%%%%%%%%%%%

In the scheme of Fig.~\ref{Figure7}, the accumulated phase rotation angle $\theta$ is the same for all states. 
The joint interrogation of the first two states (using the method discussed in Rec.~\ref{Sec.Joint}) provides a ``rough'' estimate, $\Theta_{AB}(\mu_A, \mu_B)$, of the true value $\theta \in [-\pi, \pi]$, 
depending on the measurement results $\mu_A$ and $\mu_B$, see Eqs. (\ref{thest12}) and (\ref{thest12b}). 
The phase feedback consists of a rotation of the spin-squeezed state around the $y$ axis by an angle $\Theta_{AB}(\mu_A, \mu_B)$.
Overall, the spin squeezed state (\ref{SSS3}) is rotates
by an angle $\theta_C(\mu_A, \mu_B) = \theta - \Theta_{AB}(\mu_A, \mu_B)$ around the $y$ axis, before the final readout.
The feedback rotation aligns the spin-squeezed state along the equator of the generalized Bloch sphere, see Fig.~\ref{Figure7}, where the state is maximally sensitive~\cite{AndrePRL2004, BorregaardRPL2013_a, BravermanNJP2018, PezzePRL2020}. The Ramsey signal for the third clock is
\be
\mean{\hat{J}_z(\theta_C)}_{\rm out} = \mean{\hat{J}_x}_{\rm in} \sin \theta_C.
\ee
A measurement of the relative number of particles in the third clock (with result $\mu_C$) leads to the estimate
\be
\Theta_{C}(\mu_C) =  {\rm arcsin} \frac{\mu_C}{\mean{\hat{J}_x}_{\rm in}}
\ee
of $\theta_C$. 
This value is added to $\Theta_{AB}(\mu_A, \mu_B)$ giving the estimate,
\be \label{thetaest123}
\Theta_{ABC}(\bm{\mu}) = \Theta_{AB}(\mu_A, \mu_B) + \Theta_{C}(\mu_C)
\ee
of $\theta$, where $\bm{\mu} \equiv \{ \mu_A, \mu_B, \mu_C\}$.

%%%%%%%%%%%%%%%%%%%%%%%%%%%%%%%%%%%%%%%%%%%%%%%
%% Figure 8
%%%%%%%%%%%%%%%%%%%%%%%%%%%%%%%%%%%%%%%%%%%%%%%
\begin{figure}[t!!]
\includegraphics[width=\columnwidth]{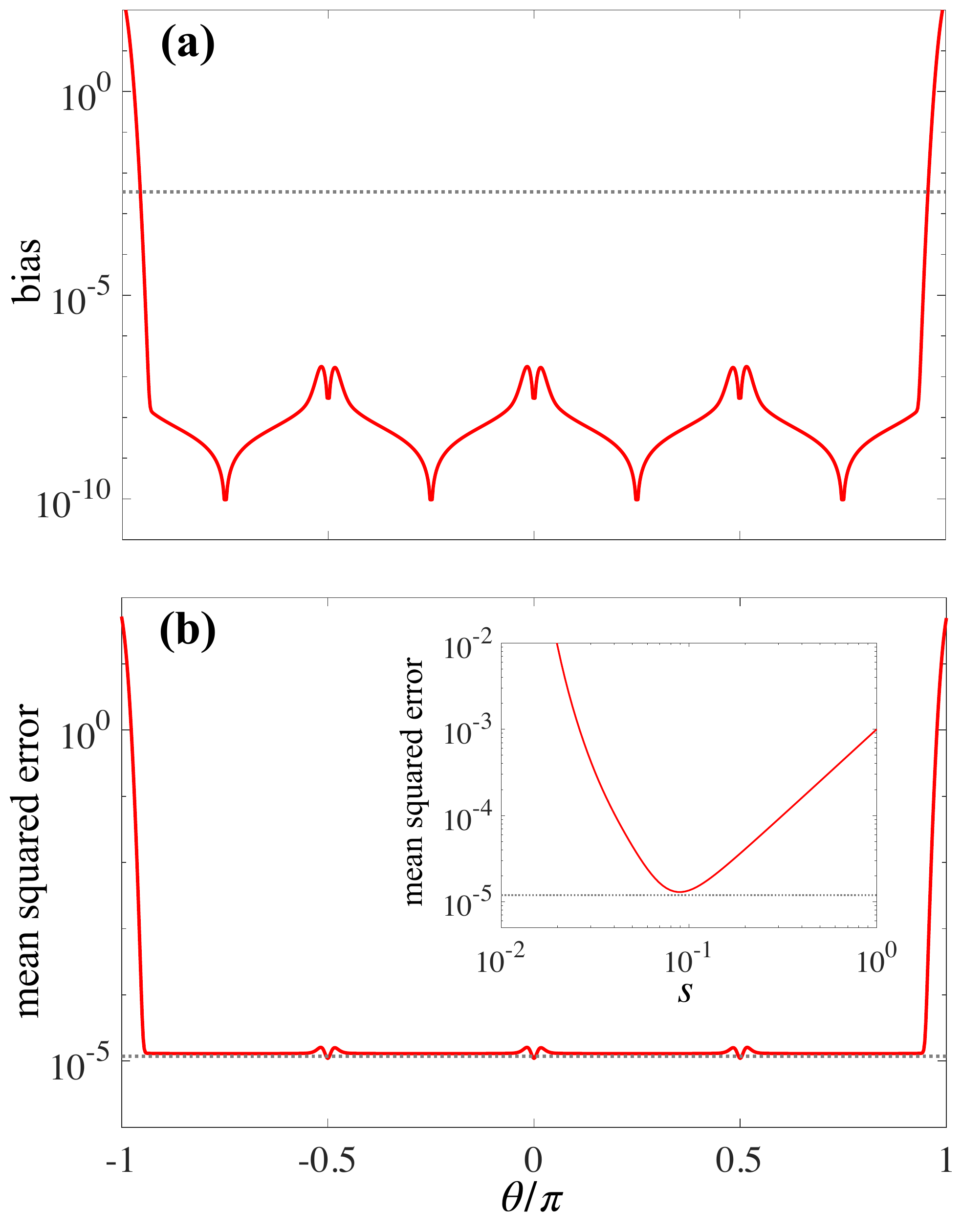}
\caption{
Bias $\vert \theta - \bar{\Theta}_{ABC}(\theta) \vert$ (a)
and mean squared error (b) as a function of $\theta$ (solid red lines). 
Results are obtained for the squeezing parameter $s^2_{\rm opt} = 1/(2N^2)^{1/3}$ and $N=1000$.
The inset shows Eq.~(\ref{EC2}) as a function of the squeezing parameter. 
In panel (b) and in the inset, the dotted line is Eq.~(\ref{EC3}).  
In panel (a), the dotted line is the square root of Eq.~(\ref{EC3}).
}
\label{Figure8}
\end{figure}
%%%%%%%%%%%%%%%%%%%%%%%%%%%%%%%%%%%%%%%%%%%%%%%
%%%%%%%%%%%%%%%%%%%%%%%%%%%%%%%%%%%%%%%%%%%%%%%
%%%%%%%%%%%%%%%%%%%%%%%%%%%%%%%%%%%%%%%%%%%%%%%

%%%%%%%%%%%%%%%%%%%%%%%%%%%%%%%%%%%%%%%%%%%%%%%
%% Figure 9
%%%%%%%%%%%%%%%%%%%%%%%%%%%%%%%%%%%%%%%%%%%%%%%
\begin{figure*}[t!!]
\includegraphics[width=\textwidth]{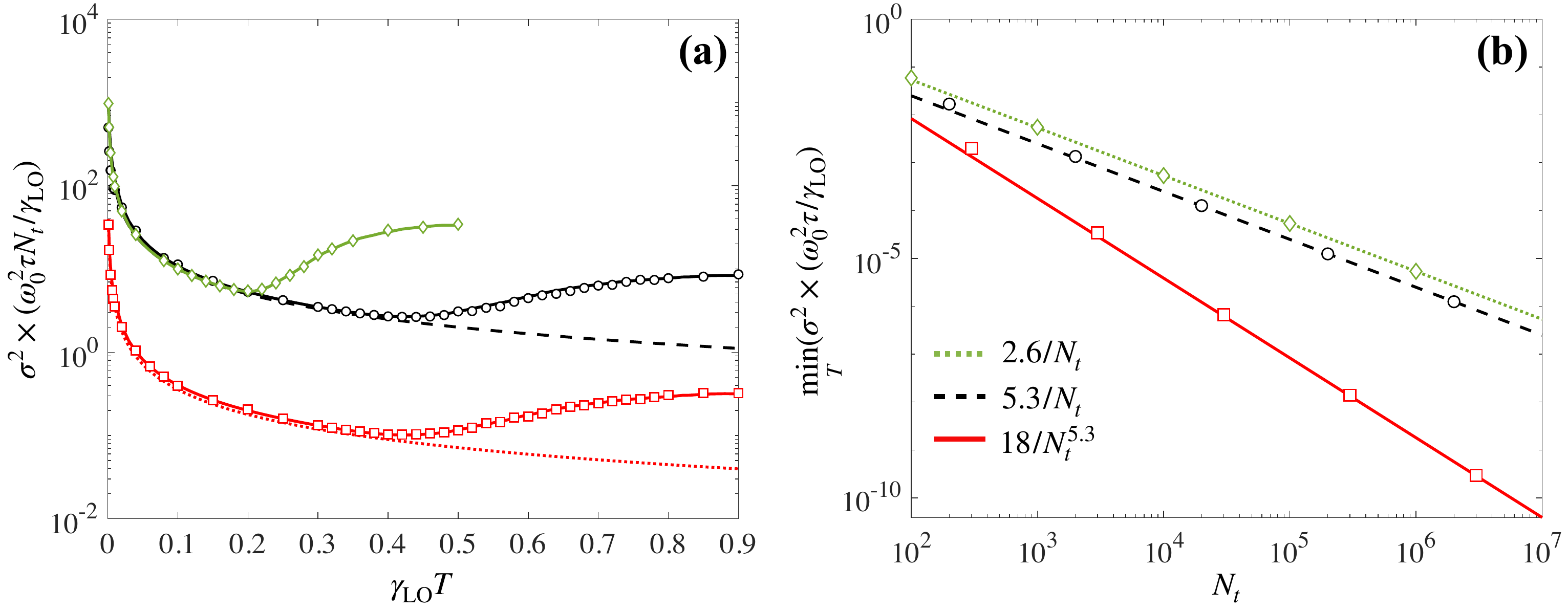}
\caption{
Panel (a) shows the Allan variance for:
i) a single Ramsey clock using a coherent state and $N_t =3000$ atoms (green diamonds); 
ii) a joint Ramsey clocks using two coherent spin states with $N_t/2 =1500$ atoms each (black circles);
iii) a hybrid clock using two coherent spin states and one optimized spin-squeezed state, each with $N_t/3 = 1000$ atoms each (red squares).
Symbols are results of ab-initio numerical simulations, while the solid lines are semi-analytical predictions (see text).
The dashed line is $\sigma^2_{\rm SQL} = 1/(\omega_0^2 T \tau N_t)$.
The dotted line is Eq.~(\ref{sigma_subSQL}).
(b) Scaling of the Allan variance (minimized over the Ramsey time) as a function of $N_t$ for the different clock protocols:
i) the single-Ramsey (green diamonds);
ii) the joint-Ramsey (black circles);
iii) the hybrid-Ramsey (red squares).
Lines are fits: the dashed and dotted lines are $\sigma^2 =O( 1/N_t)$, while the solid line is $\sigma^2 =O( 1/N_t^{5/3})$.
The numerical results in both panels are obtained for flicker LO noise and $\gamma_{\rm LO} \tau = 100$.
}
\label{Figure9}
\end{figure*}
%%%%%%%%%%%%%%%%%%%%%%%%%%%%%%%%%%%%%%%%%%%%%%%
%%%%%%%%%%%%%%%%%%%%%%%%%%%%%%%%%%%%%%%%%%%%%%%
%%%%%%%%%%%%%%%%%%%%%%%%%%%%%%%%%%%%%%%%%%%%%%%

In Fig.~\ref{Figure8}(a) we plot the bias, 
$\vert \theta - \bar{\Theta}_{ABC}(\theta)\vert$ as a function of $\theta$, where $\bar{\Theta}_{ABC}(\theta) = \mathcal{E}_{\bm{\mu} \vert \theta} [\Theta_{ABC}(\bm{\mu}]$, 
$\mathcal{E}_{\bm{\mu} \vert \theta}[...] = \sum_{\bm{\mu}}\Theta_{ABC}(\bm{\mu}) P(\bm{\mu} \vert \theta) ...$ indicates the statistical averaging, and
$P(\bm{\mu} \vert \theta) = P(\mu_A\vert \theta) P(\mu_B \vert \theta) P(\mu_C \vert \theta - \Theta_{AB}(\mu_A, \mu_B))$.
In Fig.~\ref{Figure8}(b) we plot
mean squared error, $\mathcal{E}_{\bm{\mu} \vert \theta} [(\Theta_{ABC}(\bm{\mu}) -\theta)^2]$, 
of $\Theta_{ABC}$.
Taking into account Eq.~(\ref{thetaest123}),
we obtain
\beq \label{EC1}
\mathcal{E}_{\bm{\mu} \vert \theta} [(\Theta_{ABC}(\bm{\mu}) -\theta)^2] &=& \sum_{\mu_A, \mu_B} P(\mu_A\vert \theta) P(\mu_B \vert \theta) \times \nonumber \\ 
&&\qquad  \times \mathcal{E}_{\mu_C \vert \theta} \Big[\big(\Theta_{C}(\mu_C) -\theta_C(\mu_A, \mu_B)\big)^2\Big]. \nonumber \\
\eeq
The rotation angle $\theta_C$ is a stochastic variable, depending on the measurement results $\mu_A$ and $\mu_B$.
We can thus introduce the distribution $P(\theta_C \vert \theta) = \sum_{\mu_A, \mu_B} P(\mu_A\vert \theta) P(\mu_B \vert \theta) \delta(\theta_C - \theta_C(\mu_A, \mu_B))$, 
where $\delta$ is the Dirac delta function, and write
\be \label{MSEb}
\mathcal{E}_{\bm{\mu} \vert \theta} [(\Theta_{ABC}(\bm{\mu}) -\theta)^2] = \int d \theta_C P(\theta_C \vert \theta) \mathcal{E}_{\mu_C \vert \theta_C} \Big[\big(\Theta_{C}(\mu_C) -\theta_C\big)^2\Big],
\ee
The mean squared error of $\Theta_{ABC}$ is thus given by 
a weighted average of the mean squared error of $\Theta_C$.
Equation (\ref{MSEb}) can be reduced thanks to the spin-squeezing of the state $\ket{\psi_C}$, provided that 
$P(\theta_C \vert \theta)$ is sufficiently narrow.
To quantify this effect we first write 
$\mathcal{E}_{\mu_C \vert \theta_C} \big[ \big( \Theta_C(\mu_C) - \theta_C \big)^2 \big]= (\Delta  \Theta_C)^2 + \big(\theta_C - \bar{\Theta}_C(\theta_C) \big)^2$, 
where $\bar{\Theta}_C(\theta_C)$ and $(\Delta  \Theta_C)^2$ are, respectively, the statistical mean value and variance of the estimator $\Theta_C$.
We then calculate the estimator variance $(\Delta \Theta_{C})^2$ via error propagation, obtaining
\be \label{theta3Errorporp}
\big(\Delta \Theta_{C} \big)^2 \approx \frac{\big(\Delta \hat{J}_z(\theta_C)\big)^2_{\rm out}}{(d\mean{\hat{J}_z(\theta_C)}_{\rm out})/d\theta_C)^2} = 
\frac{(\Delta \hat{J}_y)^2_{\rm in}}{\mean{\hat{J}_x}^2_{\rm in}} + \frac{(\Delta \hat{J}_x)^2_{\rm in}}{\mean{\hat{J}_x}^2_{\rm in}}  \tan^2 \theta_C.
\ee
Neglecting the bias, namely taking $\vert \theta_C - \bar{\Theta}_C(\theta_C) \vert \ll \Delta  \Theta_C$, which is verified numerically, see Fig. \ref{Figure8}(a),
we obtain 
\be \label{EC}
\mathcal{E}_{\bm{\mu} \vert \theta} \big[(\Theta_{ABC}(\bm{\mu}) -\theta)^2\big] =
\frac{(\Delta \hat{J}_y)^2_{\rm in}}{\mean{\hat{J}_x}^2_{\rm in}} + \frac{(\Delta \hat{J}_x)^2_{\rm in}}{\mean{\hat{J}_x}^2_{\rm in}}  \int d\theta_C\,P(\theta_C \vert \theta)\,\theta_C^2,
\ee
where we have used $\tan^2  \theta_C   \approx \theta_C^2$.
Finally, $P(\theta_C \vert \theta)$ can be taken as a Gaussian distribution with width given by the mean squared error of $\Theta_{AB}$, 
Eq.~(\ref{fluctuationsAB}).
As discussed above, Eq.~(\ref{fluctuationsAB}) is essentially constant and given by $1/(2N)$ with a slight dependence on $\theta$ around $0, \pm \pi/2$, as shown by the black line in Fig.~\ref{Figure2}(b).
Replacing $P(\theta_C \vert \theta)=  \sqrt{\tfrac{N}{\pi}} e^{-N \theta_C^2}$ into Eq.~(\ref{EC}), we thus obtain
\be  \label{EC2}
\mathcal{E}_{\bm{\mu} \vert \theta} [(\Theta_{ABC}(\bm{\mu}) -\theta)^2] =
 \frac{(\Delta \hat{J}_y)^2_{\rm in}}{\mean{\hat{J}_x}^2_{\rm in}} + \frac{(\Delta \hat{J}_x)^2_{\rm in}}{\mean{\hat{J}_x}^2_{\rm in}}  \frac{1}{2N}.
\ee
Notice that the average spin moments and variances of the state (\ref{SSS3}) be calculated analytically for $N \gg 1$ and $s^2 N \gtrsim 1$~\cite{PezzePRL2020}:
 $\mean{\hat{J}_x}_{\rm in} = (N/2) e^{-1/(2s^2N)}$, $(\Delta \hat{J}_x)^2_{\rm in} = (N^2/8)(1-e^{-1/(s^2N)})^2$.
These analytical expressions can be replaced into Eq.~(\ref{EC2}), giving 
\be
\mathcal{E}_{\bm{\mu} \vert \theta} [(\Theta_{ABC}(\bm{\mu}) -\theta)^2] =
\frac{4s^2 + (1-e^{-1/(s^2N)})^2}{4 N e^{-1/(2s^2N)}}
\ee
A minimization as function of the squeezing parameter $s$ gives 
 \be \label{EC3}
 \min_s 
  \mathcal{E}_{\bm{\mu} \vert \theta} [(\Theta_{ABC}(\bm{\mu}) -\theta)^2]
\approx 
\frac{3}{2^{4/3}} \frac{1}{N^{5/3}},
 \ee
for the optimal value $s_{\rm opt}^2 = 1/(2N^2)^{1/3}$,
where the last equality holds under the condition $s^2N \gg 1$ and is obtained by keeping the leading orders in the Taylor expansion of $e^{-s^2N}$.
A plot of Eq.~(\ref{EC2}) as a function of the squeezing parameter is shown in the inset of Fig.~\ref{Figure8}(b).
The existence of an optimal squeezing parameter is a direct consequence of the bending of squeezed state in the Bloch sphere.
The bending, quantified by $(\Delta \hat{J}_x)^2_{\rm in}$, increases the output relative number of particles fluctuations for relatively large $\theta_C$, 
when compared to the single coherent spin state case [which has $(\Delta \hat{J}_x)^2\textbf{}=0$]. 
This increase of measurement uncertainty corresponds to an increase of phase uncertainty $\Delta \Theta_C$, according to Eq.~(\ref{theta3Errorporp}). 
The state bends more and more in the Bloch sphere as $s$ decreases. 
The dashed line in both the main panel and in the inset of Fig.~\ref{Figure8}(b) is Eq.~(\ref{EC3}): the agreement with the numerical calculation of the mean square error
is excellent apart the expected wiggles around $\theta=0$ and $\pm \pi/2$, and the increase close to $\pm \pi$ due to the strong bias of $\Theta_{ABC}$
(which, in turn, is due to the bias of $\Theta_{AB}$, as discussed in Sec.~\ref{Sec2}).

Figure~\ref{Figure9}(a) shows the Allan variance as a function of the interrogation time for three strategies: 
i) a single-Ramsey clock in a coherent spin state of $N=3000$ atoms (green diamonds);
ii) the joint-Ramsey clock strategy using two coherent spin states of $N=1500$ atoms each (black circles); and 
iii) the approach combining joint-Ramsey interrogation and spin squeezing (red squares), using three states of $N=1000$ atoms each, maximized over $s$.
Symbols are results of ab initio numerical simulation of the Ramsey scheme (without using any of the theoretical assumptions considered above).
The solid lines are the semi-analytical prediction of Eq. (\ref{AllanFinal}).
In particular, 
The dotted red line is the analytical prediction
\be \label{sigma_subSQL}
\sigma^2 = \bigg( \frac{9}{2} \bigg)^{4/3} \frac{1}{\omega_0^2 T \tau  N^{5/3}},
\ee
that is obtained by neglecting phase slip effects, where $N_t = 3N$. 
In Fig.~\ref{Figure8}(d) we show the scaling of the optimal Allan variance (minimized over Ramsey time $T$) as a function of the number of particles. 
The solid line is the numerical fit that confirms the predicted $\sigma^2 = O(1/N_t^{5/3})$ behaviour.
Furthermore, following Ref.~\cite{PezzePRL2020} it is possible to extend the scheme of Fig.~\ref{Figure7} to a cascade of $k$ squeezed states with decreasing optimal squeezing parameter $s$ and reach a scaling of absolute clock stability
$\sigma^2 =O( N_t^{-2+1/3^{k}})$. 
The optimization of the number of particles in each ensemble~\cite{PezzeARXIV} leads to analogous scalings but with improved prefactors.  
\section{Conclusions}
\label{Sec6}

In this manuscript we have proposed a joint-Ramsey interrogation method where two coherent spin states are prepared ``out-of-phase'', 
namely pointing along the $x$ and $y$ axis of the generalized Bloch sphere, respectively, and interrogate the same LO. 
The joint interrogation allows to extend the inversion region for the unbiased estimation of a collective rotation angle $\theta$ from $[-\pi/2,\pi/2]$ (that is the case of a single Ramsey clock) to $[-\pi,\pi]$. 
This effectively extends the optimal Ramsey interrogation time (identified as the minimum of the Allan variance) and thus increases the absolute stability.
We have demonstrated an improvement in the long-term stability of a factor 2 for $1/f$ (flicker) noise and of a factor 4 for white noise.
The joint interrogation method is reminiscent of a protocol first introduced by Kitaev in the context of quantum phase estimation with single qubits~\cite{Kitaev}.
The idea is here extended to coherent spin states of a large number of qubits and adapted in the context of atomic clocks.

It is important to clarify the significance of the figure of merit considered in this work.
We have calculated the Allan variance of the stochastic variable Eq.~(\ref{yAllan}), given by the difference between 
the true value of the accumulated phase and the estimated one at each Ramsey interrogation. 
Furthermore, the interrogation is stopped when the true value of the phase is found outside the inversion region.
Experimentally, the true value of the phase is inaccessible. 
The significance of our study is to give a ``safe'' maximum Ramsey time where phase slips are negligible. 
Our paper thus shows that this safe maximum interrogation time can be extended when using the joint interrogation method.
Finally, we have shown that the joint protocol can be combined with a recent proposal using spin-squeezed states to obtain a scaling of the stability
faster than the SQL $\sigma^2_{\rm SQL} = O(1/N_t)$.
Our proposal can be readily realized in state-of-the-art experimental implementations and addresses one of the  major problems for current atomic clocks.

\acknowledgments{
We thank F. Levi, K. Hammerer, M. Schulte and M. Tarallo for discussions. 
We acknowledge funding of the project EMPIR-USOQS, EMPIR projects are co-funded by the European Union’s Horizon2020 research and 
innovation program and the EMPIR Participating States.
We also acknowledge support by the H2020 QuantERA ERA-NET cofund QCLOCKS.
This research was supported by the 111 project (Grant No. D18001),
the Hundred Talent Program of the Shanxi Province (2018).}

\begin{center}
   {\bf APPENDIX} 
\end{center}

\begin{figure}[h!]
\includegraphics[width=\columnwidth]{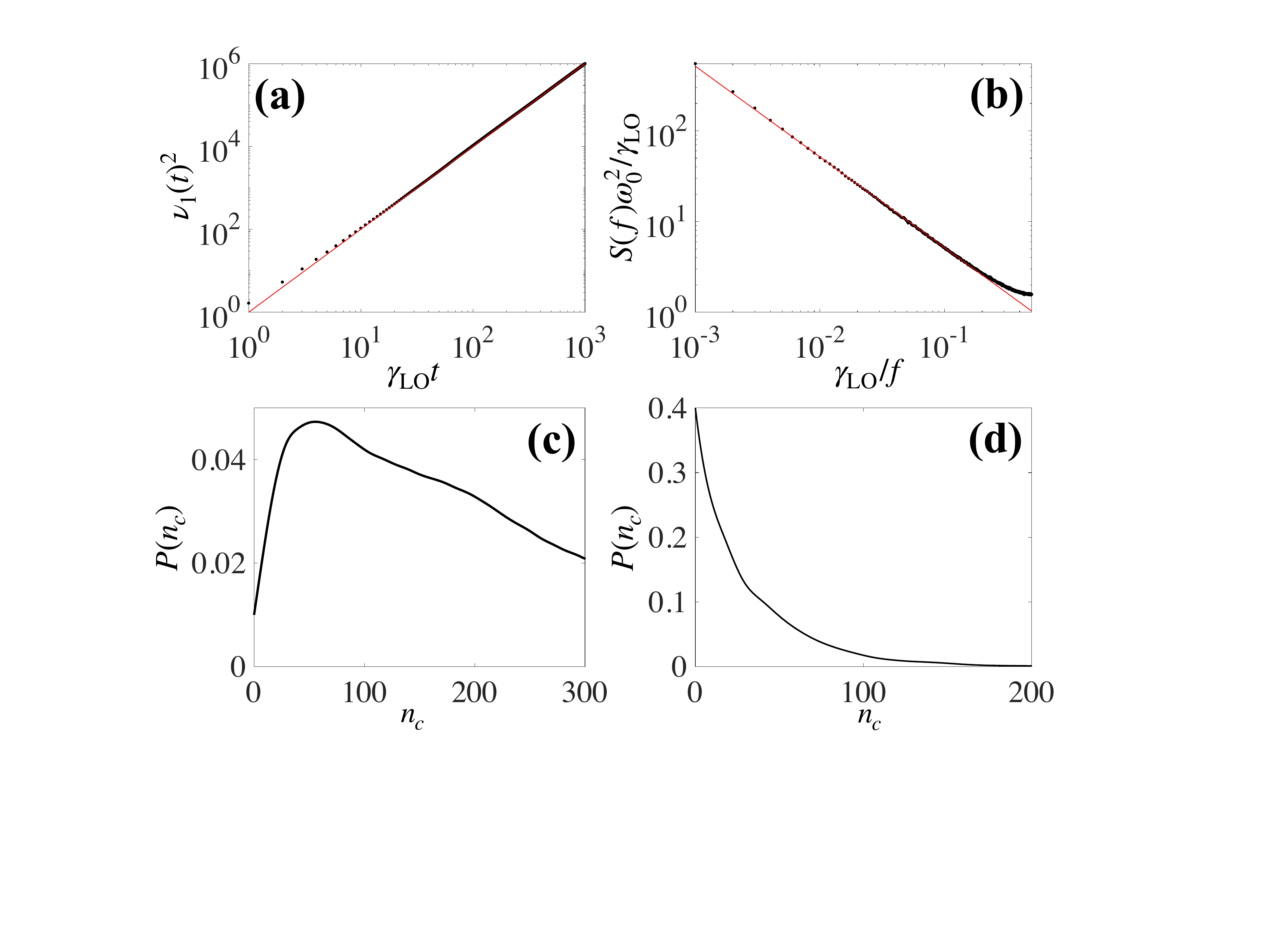}
\caption{(a) Phase variance as a function of $\gamma_{\rm LO}t$ (dots). The solid line is $v_1(t)^2 = (\gamma_{\rm LO} t)^2$. 
(b) Power spectral density as a function of 
$\gamma_{\rm LO}t$ (dots). The solid line is a fit 
$S(f) = h_{\rm LO}/f$ giving $h_{\rm LO} = 1/(2 \chi \log 2) \times (\gamma_{\rm LO}/\omega_0)^2$ with $\chi = 1.4$.
Panels (c) and (d) show $P_T(n_c)$ for $\gamma_{\rm LO}T=0.3$ and $\gamma_{\rm LO}T=0.45$, respectively.}
\label{Figure10}
\end{figure}

We give here details on the numerical simulations performed in this manuscript. 
We numerically generate the correlated-noise LO signal following the discrete incremental method outlined in Ref.~\cite{KasdinIEEE1995}.
In Fig.~\ref{Figure10} we show results of numerical simulations of the flicker noise generated numerically. 
The numerical code generates a vector $\delta \tilde{\omega}_{\rm LO}(t)/\omega_0$ at discrete times.
In panel (a) we show the phase variance $v_1(t)^2 = \mathcal{E}_{\tilde{\omega}}[\tilde{\theta}_{\rm LO}(t)^2]$, 
where ${\mathcal{E}}_{\tilde{\omega}}$ indicates statistical averaging over LO fluctuations (obtained from $10^4$ numerical realizations, the vector size being $n = 10^3$) 
and 
$\tilde{\theta}(t) = \int_{0}^t d\tilde{t} \, \tilde{\omega}_{\rm LO}(\tilde{t})$.
In panel (b) we plot the power spectral density 
$S(f) = \mathcal{E}_{\tilde{\omega}} [ \vert \delta \tilde{\omega}_{\rm LO}(f) \vert^2 ]$, where 
$\delta \tilde{\omega}_{\rm LO}(f)$ is the Fourier transform of the noise signal $\delta \tilde{\omega}_{\rm LO}(t)$.
The numerical results follows very well 
the expected behaviour  $v_1(t)^2 = (\gamma_{\rm LO} t)^2$ and $S(f) \sim 1/f$ except for short times and large frequencies where there are some deviations.

In panels (c) and (d) of Fig.~\ref{Figure10} we show examples of $P_T(n_c)$ for flicker noise, obtained for $\gamma_{\rm LO} T=0.3$ and $\gamma_{\rm LO} T=0.45$, respectively.
Here and in the numerics shown in the main text, $P_T(n_c)$ are calculated from $5 \times 10^4$ noise realizations. 

In the numerical simulations of the clock protocols the time step of the numerical noise generation is set equal to the Ramsey time $T$. In the simulations including dead time, the time step is set to $T_D$ and $T$ is taken as a multiple of $T_D$. 
Measurement results are generated numerically using the exact $P(\mu \vert \theta) = \vert \bra{\mu} \hat{U}(\theta) \ket{\psi} \vert^2$, depending on the interferometer input $\ket{\psi}$ and transformation $\hat{U}(\theta)$. 
For large number of particles $N$, $P(\mu \vert \theta)$ is replaced by a Gaussian distribution centered at $\mean{\hat{J}_z(\theta)}_{\rm out}$ and of width 
$\big(\Delta \hat{J}_z(\theta)\big)_{\rm out}^2$:
the two approaches give the same results for small $N$.
Statistical averaging in all plots shown in this manuscript is typically obtained for $2 \times 10^4$ realizations.

%%%%%%%%%%%%%%%%%%%%%%%%%%%%%%%%%%%%%%%%%%%%%%%
%% Figure 10
%%%%%%%%%%%%%%%%%%%%%%%%%%%%%%%%%%%%%%%%%%%%%%%
%%%%%%%%%%%%%%%%%%%%%%%%%%%%%%%%%%%%%%%%%%%%%%%
%%%%%%%%%%%%%%%%%%%%%%%%%%%%%%%%%%%%%%%%%%%%%%%
%%%%%%%%%%%%%%%%%%%%%%%%%%%%%%%%%%%%%%%%%%%%%%%

%


\begin{thebibliography}{10}
\bibitem{nota3}
Unless when necessary, we drop here the suffix A and AB to indicate the estimate obtained from the single and joint Ramsey interferometers, respectively.
We also indicate with $\mu$ the result of a measurement, which is a single value for the single Ramsey interferometer or a double value in the case of the joint interferometer.
\bibitem{noteAllan}
It should be noticed that, 
if no phase slip occurs, 
the quantity $y_{n}$ is typically small, of the order of $1/\sqrt{N}$. Instead, if one or more phase slips occur up to the $n$th Ramsey cycles, then the quantity $y_{n}$ is typically much larger than $1/\sqrt{N}$. 
Relatively rare slip events thus dominate the calculation of the Allan variance making it statistically unstable.
\end{thebibliography}
\end{document}